\documentclass[%
 reprint,
 amsmath,amssymb,
 aps,
 nofootinbib,
]{revtex4-1}

\usepackage{graphicx}% Include figure files
\usepackage{dcolumn}% Align table columns on decimal point
\usepackage{bm}% bold math
\usepackage{titlesec}
\usepackage{alphalph}
\usepackage{hyperref}

\begin{document}

\title{Physical limits to biomechanical sensing}% Force line breaks with \\

\author{Farzan Beroz,$^{1,2}$ Louise M. Jawerth,$^{3,4}$ Stefan M\"unster,$^{3,5}$ \\ David A. Weitz,$^{4,5}$ Chase P. Broedersz,$^{1,2,6,a}$ and Ned S. Wingreen$^{7,b}$ \\
 $\mathit{^1}$\textit{Joseph Henry Laboratories of Physics, Princeton University, Princeton NJ 08544, USA}, \\
 $\mathit{^2}$\textit{Arnold-Sommerfeld-Center for Theoretical Physics and Center for NanoScience, Ludwig-Maximilians-Universit\"at
    		M\"unchen, Theresienstrasse 37, D-80333 M\"unchen, Germany},\\
 $\mathit{^3}$\textit{Department of Biological Physics, Max Planck Institute for the Physics of Complex Systems, N\"othnitzer Str. 38, 01187 Dresden, Germany},\\
 $\mathit{^4}$\textit{Department of Physics, Harvard University, Cambridge, MA 02138},\\
 $\mathit{^5}$\textit{School of Engineering and Applied Sciences, Harvard University, Cambridge, MA 02138},\\
 $\mathit{^6}$\textit{Lewis-Sigler Institute for Integrative Genomics, Princeton University, Princeton NJ 08544, USA},\\
 $\mathit{^7}$\textit{Department of Molecular Biology, Princeton University, Princeton NJ 08544, USA}\\
}

%\affil[1]{Department of Computer Science, \LaTeX\ University}
%\affil[2]{Department of Mechanical Engineering, \LaTeX\ University}

\date{\today}% It is always \today, today,

\begin{abstract}
Cells actively probe and respond to the stiffness of their surroundings. Since mechanosensory cells in connective tissue are surrounded by a disordered network of biopolymers, their \emph{in vivo} mechanical environment can be extremely heterogeneous. Here, we investigate how this heterogeneity impacts mechanosensing by modeling the cell as an idealized local stiffness sensor inside a disordered fiber network. For all types of networks we study, including experimentally-imaged collagen and fibrin architectures, we find that measurements applied at different points throughout a given network yield a strikingly broad range of local stiffnesses, spanning roughly two decades. We verify via simulations and scaling arguments that this broad range of local stiffnesses is a generic property of disordered fiber networks, and show that the range can be further increased by tuning specific network features, including the presence of long fibers and the proximity to elastic transitions. These features additionally allow for a highly tunable dependence of stiffness on probe length. Finally, we show that to obtain optimal, reliable estimates of global tissue stiffness, a cell must adjust its size, shape, and position to integrate multiple stiffness measurements over extended regions of space.
\end{abstract}

\maketitle

\renewcommand*{\thefootnote}{\alph{footnote}}
\footnotetext{\href{mailto:c.broedersz@physik.uni-muenchen.de}{Electronic address: \nolinkurl{c.broedersz@physik.uni-muenchen.de}}}
\footnotetext{\href{mailto:wingreen@princeton.edu}{Electronic address: \nolinkurl{wingreen@princeton.edu}}}

Mechanical cues can govern cellular behavior in decisive ways (1-2). The elastic properties of a cell's substrate have been shown to guide cell migration (3, 4) and determine cell fate (5, 6). Eukaryotic cells, including fibroblasts, mesenchymal stem cells, and cancer cells, attach to substrates via transmembrane protein complexes called focal adhesions, allowing the cell to sense stiffness (2, 7-10). Knockdown studies have established that this mechanosensing contributes both to motility and to the regulation of cell shape in three-dimensional \emph{in vitro} systems that closely resemble \emph{in vivo} cellular environments (10-13), where cells are surrounded by a loosely connected, disordered network of protein fibers such as collagen or fibrin (14). These biopolymers form a major structural component of the extracellular matrix (ECM), which serves as the physical scaffolding within which cells live and move. While it is clear that cells actively probe these extracellular networks, it remains unclear how ECM micromechanical properties impact mechanosensing.\\
\indent Both \emph{in vitro} experiments (15-19) and theory (20-24) have demonstrated that biopolymer networks exhibit rich \emph{macroscopic} mechanical behavior, depending sensitively on network connectivity. However, because the size of a typical cell is comparable to the pore size of the ECM (25), any mechanical information must be inferred by locally probing an extremely heterogeneous material. Although a few studies have begun to characterize this \emph{microscopic} response (26, 27), a theoretical understanding of how local mechanics are determined by the surrounding heterogeneous structure is still lacking. \\
\indent How does the intrinsic heterogeneity of the ECM limit a cell's ability to learn about its \emph{global} environment from purely \emph{local} mechanical measurements? For the case of chemical sensing, consideration of the fundamental physical limits dates back to Berg and Purcell's consideration of noise due to the random arrival of diffusing particles (28-30). Here, we take a similar approach to explore the fundamental limits of mechanosensing, where in contrast to chemical signals, the cues are static in time but distributed nonuniformly in space.\\
\indent To quantify the physical limits of mechanosensing imposed by a cell's disordered environment, we investigate a simple model consisting of two components: the ECM as an elastic network that deforms in response to external forces, and the cell as an idealized measurement device that probes the stiffness of its surroundings. We found that experimentally-imaged collagen and fibrin networks and randomly-generated networks all yield a \emph{very} broad range of modeled local stiffness responses, spanning roughly two decades. We trace the origin of this broad range of stiffnesses to two intrinsic features of disordered networks: (\emph{i}) the local stiffness depends primarily on a small number of local fibers with consequently large variations, and (\emph{ii}) these proximal fibers contribute to stiffness in a highly cooperative manner. The combined effect is a large uncertainty in global stiffness inference, which we demonstrate is only further increased by the presence of long fibers, proximity to elastic transitions, and correlations among nearby measurements. Finally, we argue that to obtain accurate estimates of global ECM stiffness, cells integrate multiple stiffness measurements over extended regions of space.

\section{Results}
\label{sec:local}

Cells in connective tissue can glean mechanical information about their surroundings by pulling on the individual biopolymers of the ECM. However, on the short length scale of a typical cell, the measured mechanical response is sensitive to the intrinsic structural disorder of the ECM. To investigate the role of local mechanical disorder in a physiologically relevant system, we considered collagen networks, which form the primary structural component of the ECM (14). We prepared a sample network by reconstituting fluorescently-labeled collagen type-I monomers and imaged its three-dimensional structure (31, \emph{Supporting Information}). The network is loosely connected (with an average coordination number $z\simeq2.9$) and highly heterogeneous at the cellular scale (with an average mesh size $\xi \simeq6.5\ \mu$m). This reconstituted collagen architecture was used as an input to construct a mechanical network model where the fibers are treated as elastic beams that can bend and stretch (Fig. 1\emph{A}). For simplicity, we modeled the stretching and bending of the beams, respectively, as springs and torsional springs connecting point-like vertices with stretching modulus $\mu$ and bending modulus $\kappa$. Since biopolymers are expected to be much more pliable to bending than to stretching, we chose the stretching modulus such that $\kappa \ll \mu \xi^2$. The bending modulus $\kappa$ of the torsional springs was fitted to the experimental network using data from macroscopic rheology (\emph{Supporting Information}).

%Fig 1
\begin{figure}[t!]
\includegraphics[width=\columnwidth]{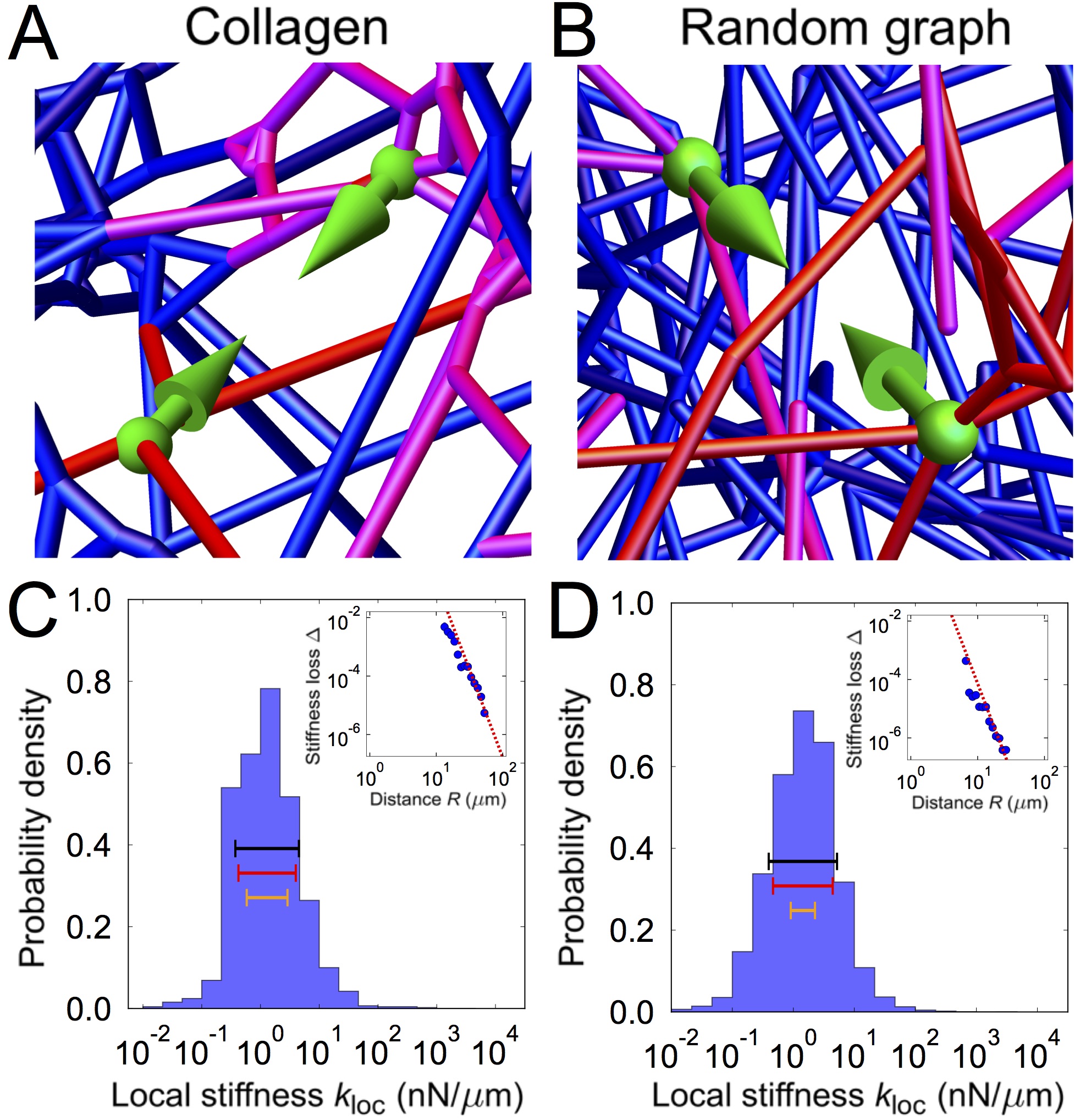}
\caption{\label{fig:F1}Force-dipole stiffness distribution.
(\emph{A} and \emph{B}) Examples of local stiffness sensing by force dipoles. Modeled deformation under stress from a local force dipole of length $d \sim 15\ \mu$m (green arrows) of (\emph{A}) experimental collagen network and (\emph{B}) RGG network. Magnitude of fiber deformations indicated by color (small deformations blue, large deformations red). (\emph{C} and \emph{D}) Distribution of local stiffnesses $k_{\mathrm{loc}}$, defined as the linear response of local deformation to a force dipole of length $d \sim 15\ \mu$m for (\emph{C}) collagen network and (\emph{D}) RGG network. Geometric standard deviation of local stiffness $\sigma_{\mathrm{loc}}$ indicated by bars (actual distribution black, estimated distribution assuming strong locality red, estimated distribution assuming weak locality orange; see \emph{Supporting Information} for details). \emph{Insets}: stiffness loss $\Delta$, defined as the relative change in local stiffness $k_{\mathrm{loc}}$ upon perturbing a network by removing a single fiber, versus distance $R$ of center of removed fiber from probe center. For collagen, probe length $d<10\ \mu$m and removed-fiber length $\ell_{ij}<10\ \mu$m, and for RGG, probe length $d<5\ \mu$m and removed-fiber length $\ell_{ij}<5\ \mu$m. Error bars are smaller than the size of data points. Dashed lines show asymptotic scaling from continuum theory, which predicts $\Delta \sim 1/R^{2D}$ for $R >> d$ (see Methods for details).}
\end{figure}

\textsf { \textbf{ Extracellular networks exhibit broad distributions of local stiffnesses spanning two decades.}} To study the mechanical response of the network to local forces that might be applied by cells,
we defined a ``local stiffness'' $k_{\textrm{loc}}$ as the linear displacement of two vertices in response to a dipole contractile force along the direction between the vertices (32). We then calculated this local stiffness
by numerically solving the equations of force balance for the network.
Strikingly, we found that local stiffness measurements yield
a very broad range of values, spanning up to roughly two decades (Fig.
1\emph{C}). Because of the large range of stiffnesses,
in what follows we characterize this variability in terms of the geometric standard deviation (Eq. S15), which we find for this collagen network to be $\sigma_{\mathrm{loc}}=0.54$.

To explore the generality of this broad local stiffness distribution, we next considered a reconstituted fibrin network (Fig. S1\emph{E}), as fibrin constitutes the main structural component of blood clots (14). We imaged a fibrin network prepared from a solution of fibrinogen and observed an average coordination number $z\simeq2.7$ and an average mesh size $\xi\simeq6.7\ \mu$m (\emph{Supporting Information}). We found that the fibrin network also has a broad distribution of local stiffnesses, with $\sigma_{\mathrm{loc}}=0.63$ very similar to that of the collagen network (Fig. S1\emph{F}). The broad distribution of local stiffnesses for both the collagen and fibrin architectures suggests that this feature may be a general property of biological fiber networks.

%Fig 3
\begin{figure}[t!]
\includegraphics[width=\columnwidth]{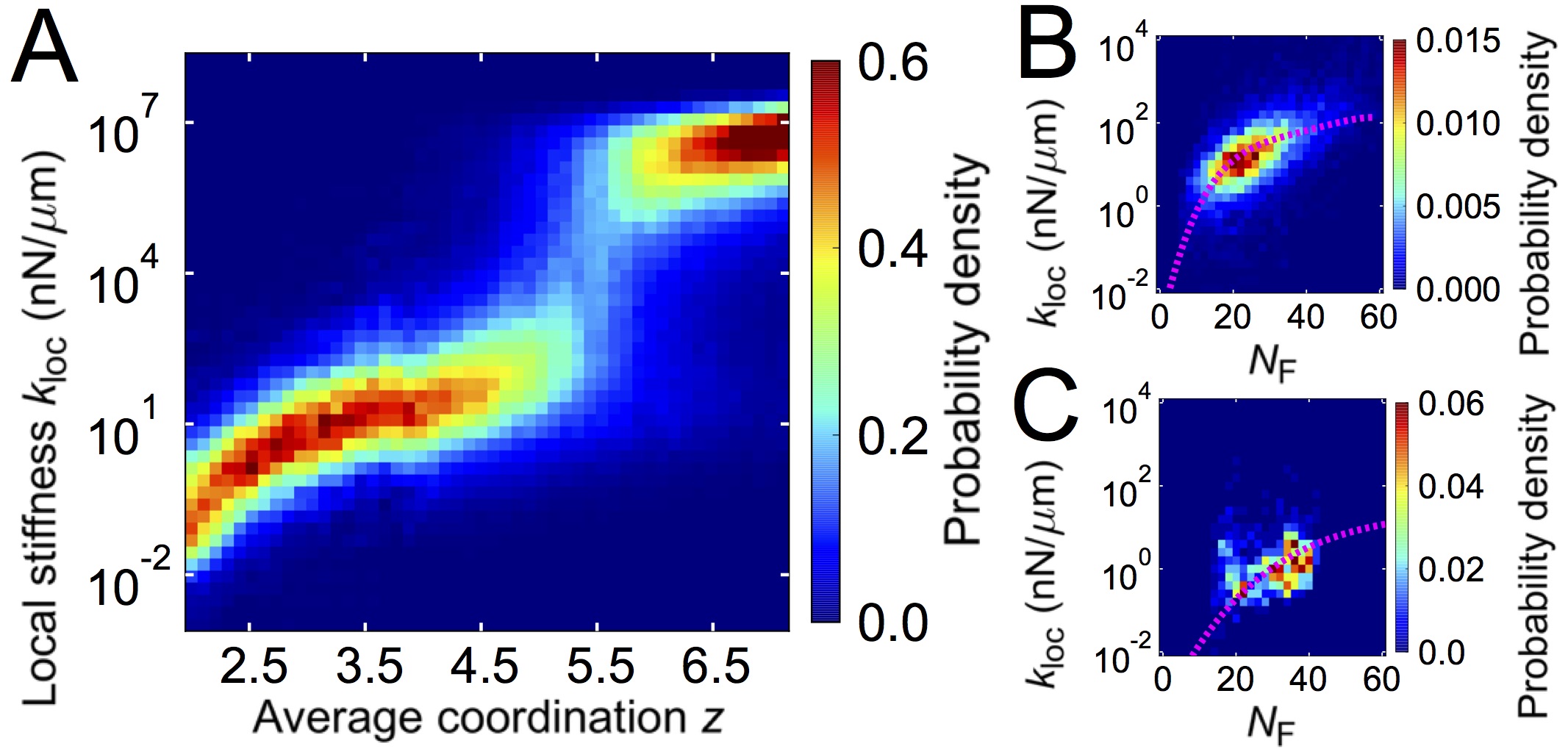}
\caption{\label{fig:F2}
Collective effect of network structure on local stiffness. (\emph{A}) Distribution of local stiffnesses $k_{\mathrm{loc}}$ for RGG network versus average coordination number of vertices, $z$ (force dipole length $d=15\ \mu$m, $\kappa / \mu = 10^{-5}\ \mu$m$^2$). (\emph{B} and \emph{C}) Joint distribution of local stiffness and number of local fibers $N_F$, defined as a weighted fraction of local bonds, with bond weight $=1$ below a short-range cutoff $\tilde{\xi}$ and decaying as $1/R^{2D}$ beyond, where (\emph{B}) $\tilde{\xi}=1.5\xi$ for RGG network ($\xi \simeq6.7\ \mu$m) and (\emph{C}) $\tilde{\xi}=2\xi$ for collagen network ($\xi \simeq 6.5\ \mu$m). Dashed lines show the macroscopic shear moduli of the networks as a function of average number of local fibers $\langle N_F \rangle$.
}
\end{figure}

To investigate the physical origins of the broad local stiffness distributions described above, we turned to idealized model networks. One way to generate model disordered networks consists of arranging vertices to lie on a regular lattice, such as a simple-cubic lattice (SC) or face-centered-cubic
lattice (FCC) in three-dimensions (Fig. S2\emph{A}, 22). Connecting these vertices randomly with a fixed probability results in a disordered lattice network. Such lattice networks have provided a useful starting point for characterizing the mechanical response
of fiber networks in two-dimensions (27). Interestingly, although three-dimensional SC and FCC
networks also yield a broad range of local stiffnesses (Fig. S4\emph{F}, Fig. S2\emph{B}),
the geometric standard deviations of these distributions ($\sigma_{\mathrm{loc}}=0.29$ and $0.37$, respectively) are considerably smaller
than those of the experimental networks (Fig. 1\emph{C} and Fig. S1\emph{F}). A possible reason for this discrepancy is that the disordered lattice networks fail to capture important structural features of real networks at the scale of the cell, such as the random positions of the vertices.

To generate a model network that better matches the structural features
of the experimental networks, we begin by distributing vertices randomly throughout a volume. The density of these vertices is set roughly equal to those of the experimental networks. Pairs of vertices are then connected according to a probability function that depends on the inter-vertex distance. This probability function was chosen to produce an average coordination number and distribution of fiber lengths that match those of the experimental architectures (\emph{Supporting Information}). The resulting modeled networks are referred
to as ``Random Geometric Graphs'' (RGG, Fig. 1\emph{B}). We computed the local stiffness
distribution for the RGG network and found that the geometric standard deviation is $\sigma_{\mathrm{loc}}=0.56$, which is approximately equal to those of the experimental networks (Fig. 1\emph{D}). Thus, in contrast to the lattice networks, the RGG network appears to quantitatively capture this important local mechanical property of the experimental networks. \newline

\textsf { \textbf{The experimental collagen and fibrin networks are in a bending-dominated regime that lies between two macroscopic elastic transitions.} }What features of the network impact the very large range of local stiffnesses we found for all types of networks? Intuitively, more fiber-dense regions should be stiffer. For a fixed density of vertices, the fiber density is roughly proportional to the network connectivity, defined as the average coordination number $z$. We briefly review how the overall connectivity of a network affects its macroscopic mechanical properties (20, 22).

For high values of $z$, the bulk elastic moduli of fiber networks are dominated by the stretching of fibers. As $z$ is lowered, the macroscopic response eventually undergoes a crossover from a stretching-dominated
to a bending-dominated response, as the network can be deformed without stretching fibers but not without bending them. As $z$ is
further lowered, another elastic transition occurs when the network ultimately loses macroscopic rigidity. Over the whole range, the macroscopic response depends strongly and nonlinearly on the network connectivity, with the strongest dependence near the two elastic transitions. We expect this sensitivity of the macroscopic stiffness to overall network connectivity to be reflected in a corresponding sensitive dependence of local stiffness on local network connectivity.

To determine how the rapid scaling of the macroscopic response with connectivity manifests locally, we varied the average coordination
number $z$ of modeled networks for a fixed ratio of the bending modulus $\kappa$ to the
stretching modulus $\mu$ (with $\kappa\ll\mu\xi^2$, Fig. 2\emph{A}, Fig. S2\emph{C}). At high connectivities,
the entire local stiffness response is dominated by stretching interactions and scales with the stretching
modulus $\mu$. As the connectivity is lowered, the entire local stiffness distribution
shifts to lower values. Specifically, the median stiffness decreases rapidly and nonlinearly
with the average coordination $z$. Near the stretching-bending crossover,
the local stiffness distribution becomes bimodal with the emergence
of a subset of probes for which the measured stiffness scales with
the bending modulus $\kappa$. Below the stretching-bending crossover, the number of stretch-dominated probes becomes negligible,
and the local stiffness response enters a bending-dominated regime. Within this regime, the median stiffness resumes rapid decay.
Finally, as the connectivity is brought below the rigidity transition, an increasing fraction of measurements yields zero stiffness
as portions of the network become floppy. We thus find that the elastic transitions of the macroscopic response manifest
locally as crossovers between stretch-dominated, bending-dominated,
and zero-stiffness measurements. At these crossovers, the median stiffness decreases most rapidly, and the local stiffness distribution becomes bimodal, yielding a maximum in the geometric standard deviation.

The low connectivity of the experimental collagen and fibrin networks
suggests that they are situated in the bending-dominated regime. At these particular connectivities, the RGG network is poised approximately halfway between the elastic transitions. Here, the median stiffness scales least rapidly, and we therefore expect the proximity to the transitions to play a minimal role in governing the width of the RGG stiffness distribution. To verify this, we systematically varied the ratio of the bending modulus to the stretching modulus for these RGG networks (Fig. S1\emph{B},\emph{D}). We found that the local stiffness measured by each individual probe is nonzero and scales with the bending modulus. This implies that the experimental networks are far from both elastic crossovers, implicating a different source for the broad local stiffness distributions. \newline

\textsf { \textbf{The local stiffness depends only on proximal network structure.}} A very broad distribution of local stiffnesses appears to be a generic feature of disordered fiber networks. What is the origin of this large variance in local stiffness measurements? To investigate how a local stiffness measurement depends
on the surrounding network structure, we calculated the stiffness loss ${\Delta}$
upon removing a single fiber (\emph{Supporting Information}). Intuitively, removal of fibers that
are more proximal to the stiffness probe should have a greater effect on $k_{\mathrm{loc}}$,
which suggests that on average, the stiffness loss should decay as
a function of the distance $R$ from the probe center to the center
of the removed fiber. Indeed, for all types of networks we studied,
we found that the average stiffness loss is consistent with $1/R^{6}$ decay (Fig. 1\emph{C},\emph{D}, Fig. S1\emph{F}, and Fig. S2\emph{B}, \emph{Insets}).

This apparent universality of the scaling of the stiffness loss suggests
that the $1/R^{6}$ power-law decay should be calculable within continuum elasticity
theory. We therefore calculated the effect of removing a single fiber in the vicinity of a local stiffness probe for the case of a uniform lattice network (\emph{Supporting Information}). The defect created by removing a fiber at a distance $R$ from the probe perturbs the dipole strain field induced by the probe. We can treat this perturbation as an additional dipole strain field originating from the defect, with a magnitude proportional to the initial strain in the removed fiber. Since both this initial strain, and the consequent additional strain ``reflected'' back to the probe, decay as the strain field of a force-dipole, i.e. as $1/R^D$ (where $D$ is the dimension), the combined effect is an increase in the strain at the location of the probe $\sim 1/R^{2D}$.

The rapid $1/R^{6}$ decay of the stiffness loss due to fiber removal
in three dimensions suggests that the local stiffness is largely determined by the network structure in the immediate vicinity of the probe. Within this small local
region, all quantities are subject to large fluctuations; for example,
since a small region typically contains a small number of fibers,
the variance of fiber density will be large. The universal, rapid $1/R^{2D}$ decay implies that
the mechanically relevant local structure will always be \emph{very}
local, with consequently large fluctuations. \newline

\textsf { \textbf{ Broad stiffness distribution arises from local density fluctuations combined with strongly cooperative contributions of proximal fibers.}} To quantify the dependence of local stiffness on the surrounding network structure, we considered
the number of local
fibers $N_{F}$, defined as the sum of fibers each weighted by a $1/R^{6}$-decaying function of its distance $R$ from the probe center (see above and \emph{Supporting Information}).
We find that $k_{\mathrm{loc}}$ and $N_F$ are well correlated for all types
of networks studied, as shown in Fig. 2\emph{B},\emph{C}. Importantly, $k_{\mathrm{loc}}$ has a strong, nonlinear dependence on
$N_F$. Specifically, the center of the marginal
distribution of local stiffnesses at fixed $N_F$ increases more rapidly than linearly
with $N_F$. This indicates that local
fibers influence local stiffness in a highly cooperative manner, i.e. combining multiple fibers typically results in much larger than additive
changes to local stiffness.

To estimate the geometric standard deviation of local stiffness,
we must account for these cooperative effects among fibers. We first
note that the scaling of the \emph{median} local stiffness with $N_F$ is consistent
with that of the macroscopic shear modulus for all types of networks we studied (Fig. S4\emph{D}-\emph{F},
S5\emph{D}-\emph{F}, \emph{Insets}). This suggests that much of the broad width
of the stiffness distribution can be accounted for by the large variations in the local fiber density taken together with the strong, nonlinear dependence of the macroscopic shear modulus on overall fiber density. To test this notion, we estimated the distribution of local stiffnesses by taking $N_F$ transformed by the functional dependence of the
macroscopic shear modulus, $G(N_{F})$, where the modulus $G$ is that of
a macroscopic network with an average number of local fibers given
by $\langle N_{F} \rangle$. Upon accounting for the strong collective effects of
fiber density in this manner, we found that the geometric standard
deviation of $G(N_{F})$ provides a very good estimate for the actual, observed
geometric standard deviation of local stiffness (Fig. 1\emph{C},\emph{D}, Fig. S1\emph{F}, Fig. S2\emph{B}, Fig. S4\emph{D}-\emph{F},
and Fig. S5\emph{D}-\emph{F}). Consequently, the estimator correctly predicts
the relative differences in the geometric standard deviation for the
different types of networks, including the observation that the
stiffness distributions for the experimental and RGG networks are
much broader than for the disordered lattice networks.

Our prediction of the width of stiffness distributions is not sensitive to the particular form of the weighting function, provided the decay is rapid enough. For instance, a hard cutoff at twice the mesh size captures a majority of the broad width for the FCC network (Fig. S4\emph{E}). However, a less rapidly decaying estimator includes a larger number of fibers in the local structure, and consequently the variance in fiber density is smaller. Thus, by comparison, an estimator with weights that decay as $1/R^{3}$ yields a smaller, much
poorer estimate of the width of the stiffness distribution (Fig. 1\emph{C},\emph{D}, Fig. S1\emph{F}, Fig. S2\emph{B}, Fig. S4\emph{D}-\emph{F}, and
Fig. S5\emph{D}-\emph{F}). \newline

\textsf { \textbf{ Long fibers increase the dynamic range of local stiffness as a function of probe length.}} Above, we demonstrated how local structure impacts local stiffness for a fixed distance between probed vertices. Biologically, this distance corresponds to the scale over which cells measure stiffness. Since cells can adopt various sizes and shapes, this means that a cell might measure on different scales depending on the length of its ``arms,'' i.e. the distance between focal adhesions. We therefore consider the local stiffness as a function of the length of the force-dipole probe in our model.

For long probes, the geometric mean and geometric standard deviation
of stiffness approach constant values (Fig. 3\emph{A},\emph{B}). This asymptotic saturation
occurs because for very long probes, the measured
deformation is simply the sum of the deformations of two independent
\emph{monopole} probes, namely the deformation at a vertex in response
to a force applied at that vertex. As the probe length is
decreased, however, the measured stiffness increases by more than a decade before the probe length reaches the average mesh size, which is a much larger increase than predicted by continuum theory (\emph{Supporting Information}). Surprisingly, the measured stiffening deviates noticeably from the continuum prediction even for probe lengths that are several times larger than the mesh size.

For the experimental and RGG networks, this anomalously large fold-stiffening is present for all values of $z$ and larger near the elastic transitions (Fig. S8). In contrast, for the FCC network, the stiffness distribution has almost no dependence on probe length away from the elastic transitions (Fig. S7\emph{D}). This implies that the dramatic fold-stiffening of the experimental and RGG networks largely results from intrinsic features of these networks that are not present for the FCC network. One relevant feature that distinguishes the different types of networks is the distribution of fiber lengths. That is, the experimental and RGG networks both contain long
fibers connecting vertices several times more distant than the mesh
size, whereas for the FCC networks, the typical fiber length (given
by the number of consecutive, coaxial bonds uninterrupted by non-coaxial
crosslinking bonds) is equal to the mesh size, i.e. the lattice spacing. Indeed, a simple extension of continuum elasticity that includes structural heterogeneity on the scale of the fiber length can quantitatively capture the majority of the large fold-stiffening observed for the experimental and RGG networks (\emph{Supporting Information}). Thus, our results indicate that the polydispersity of fiber length in biopolymer networks can increase the fold-stiffening as a function of probe length. \newline

%Fig 3
\begin{figure}[t!]
\includegraphics[width=\columnwidth]{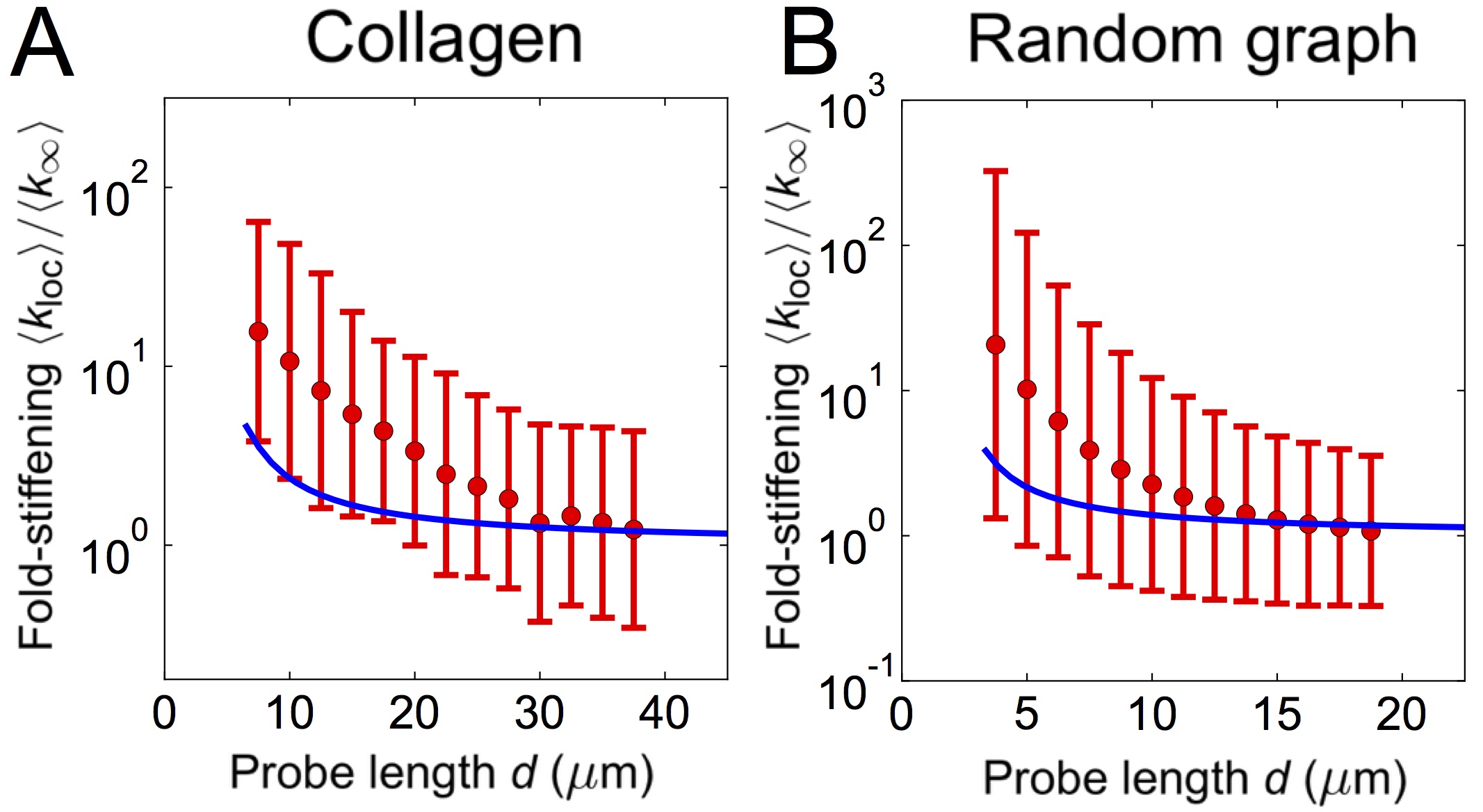}
\caption{\label{fig:F3} (\emph{A} and \emph{B}) Fold-stiffening $\langle k_{\mathrm{loc}} \rangle / \langle k_{\mathrm{\infty}} \rangle$, defined as the geometric mean stiffness $\langle k_{\mathrm{loc}} \rangle$ divided by the geometric mean stiffness for asymptotically long probes $\langle k_{\mathrm{\infty}} \rangle$ (calculated using local stiffnesses of monopole probes), versus probe length $d$ for (\emph{A}) collagen network and (\emph{B}) RGG network. Error bars show geometric standard deviation of local stiffness; solid curves (blue) from continuum theory (\emph{Supporting Information}).
}
\end{figure}

\textsf { \textbf{Size and shape of sampling region determines accuracy of stiffness inference.}} The intrinsic heterogeneity of fiber networks presents a challenge
for cells attempting to glean information from stiffness cues. That
is, tissues with different \emph{global} stiffness properties may have significant
overlap of their \emph{local} stiffness distributions. In this case, a single
local stiffness measurement would provide only a poor estimate of
the global stiffness and identity of the tissue. In view of our conclusion that local stiffness distributions are generically broad, how accurately can cells infer the global stiffness of their surroundings from samples of local stiffness?

One possible strategy for cells to increase their inference accuracy
is by averaging multiple stiffness measurements. With enough samples
of local stiffness, this method would allow cells to reliably distinguish
between global environments with different mechanical properties. We can visualize the local stiffness that a cell might infer at each
point of the network by plotting the geometric mean stiffness of measurements
obtained within a cell-sized sphere centered on that point (Fig. 4\emph{B},
\emph{Inset}). The patterns in the stiffness landscape reflect the
correlations of nearby stiffness measurements and extend over regions
larger than typical cell sizes. More precisely,
we find that the correlation function $C_{ij}$ between the log-local
stiffnesses measured by two probes decays exponentially as a function
of the distance $R_{ij}$ between their centers, with decay lengths
around 3 $\mu$m (Fig. 4\emph{B} and Fig. S10\emph{A}, \emph{Insets}). These spatial correlations of local stiffness
arise because nearby stiffness measurements depend on shared local
structure. Since correlations reduce the effective number of independent
samples, we expect less accurate global inference if stiffness measurements
are made closer together in space.

To quantitatively study the reduction in accuracy of cellular mechanosensing due to spatial correlations, we
modeled cell inference as an idealized sampling and averaging. To be concrete, we considered the sampled stiffness to be the geometric
mean of a random sample of three probes whose centers are contained
within prolate spheroids of varying volume and aspect ratio (Fig. 4\emph{A}, \emph{Inset}). For all types of networks we studied, the shape of the
sampling region directly impacts the uncertainty of stiffness inference
(defined as the geometric standard deviation $\sigma_{3}$ of the
inferred stiffness, Fig. 4 and Fig. S10\emph{B}). Specifically, the uncertainty is always reduced
by increasing the volume of the sampling region as well as by increasing
its aspect ratio. As the volume and aspect ratio of the sampling region
are increased, typical samples of stiffness become increasingly uncorrelated,
and, in the asymptotic limit, the uncertainty of the sampled stiffnesses
approaches the geometric standard deviation for independent samples.

%Fig4
\begin{figure}[t!]
\includegraphics[width=\columnwidth]{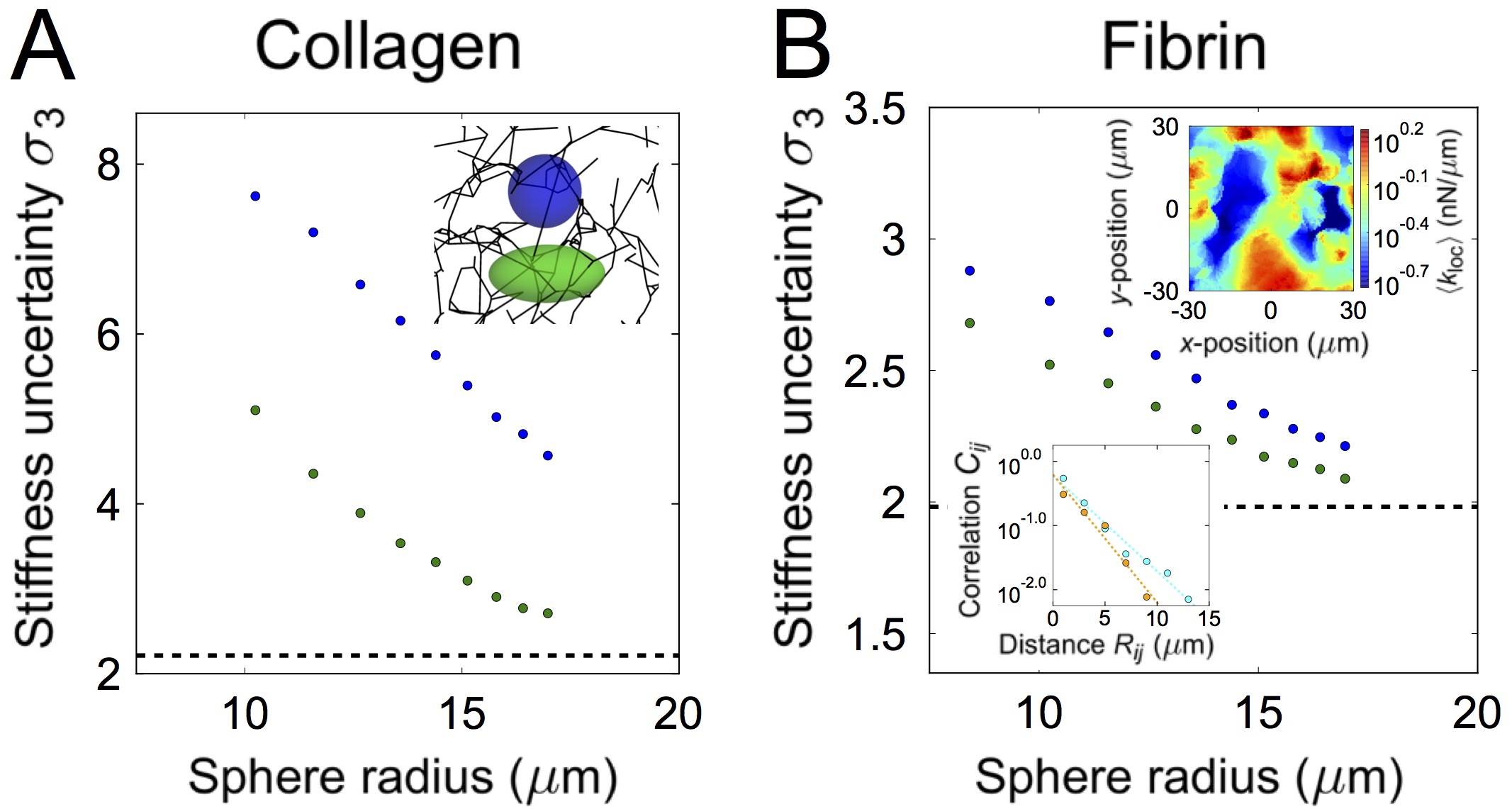}
\caption{\label{fig:F4}Minimum uncertainty of stiffness inference. (\emph{A} and \emph{B}) Geometric standard deviation $\sigma_{\mathrm{3}}$ of the geometric mean of a random sample of $N=3$ local stiffnesses measured by force dipoles whose centers lie within spheres (\emph{A}, \emph{Inset}: blue) and prolate spheroids of equivalent volume (\emph{A}, \emph{Inset}: green, aspect ratio 2:1, and red, aspect ratio 3:1) versus radius of spheres for (\emph{A}) experimental collagen network and (\emph{B}) experimental fibrin network. Dashed black lines shows the geometric standard deviation $\sigma_{\mathrm{loc}}^{1/\sqrt{3}}$ of the geometric mean of $N=3$ independent local stiffness measurements. Upper \emph{Inset} of \emph{B} shows a two-dimensional slice of the fibrin network stiffness landscape, defined at each point as the geometric mean of local stiffnesses measured by all force dipoles whose centers lie within a sphere of radius $8.4 \mu$m centered on the point. Lower \emph{Inset} of \emph{B} shows the two-point geometric correlation $\rho_{ij}$ for local stiffness, defined as the covariance between log-local stiffnesses $\log k_{\mathrm{loc}}^{(i)}$ and $\log k_{\mathrm{loc}}^{(j)}$ divided by the logarithm of the geometric standard deviation of local stiffness squared $(\log \sigma_{\mathrm{loc}})^2$, versus distance $R_{ij}$ for collagen (orange) and fibrin (cyan) networks.
}
\end{figure}

\section{Discussion}

Mechanical cues can guide cell motility and differentiation (3-6). Indeed, cells are observed to reliably distinguish among two-dimensional homogeneous substrates with bulk stiffnesses similar to brain ($\sim 1$ kPa), muscle ($\sim 10$ kPa), and bone ($\sim 100$ kPa) (5). However, the three-dimensional \emph{in vivo} mechanical environment at the cellular scale can be extremely heterogeneous (14, 25). Here, we asked how this intrinsic heterogeneity impacts mechanosensing. Interestingly, we found that within macroscopically homogeneous but locally disordered collagen and fibrin networks, local probes yield a very broad range of stiffnesses, spanning roughly two decades (similar to the relative difference in the bulk stiffness of brain and bone). Moreover, the average measured stiffness is anomalously sensitive to probe length. We quantitatively captured these striking features of the experimental networks with modeled networks, which enabled us to elucidate their physical origins using a combination of simulations and scaling arguments.

We first established that the very broad range of local stiffnesses is universal across network types and spans a wide range of connectivities, including multiple elastic regimes. We then traced the origin of this pervasive broad distribution to variations in local structure. Specifically, stiffness probes are primarily sensitive to a very small region of local fibers, and these privileged fibers contribute to the measured response in a highly cooperative fashion. Finally, we found that the distribution of stiffnesses can be further broadened by tuning specific network features, including proximity to elastic transitions and geometrical disorder. While our results indicate that the collagen and fibrin networks are poised squarely in a regime where the response is dominated by fiber bending, these networks include strong geometrical disorder, including randomly-oriented fiber junctions and a polydisperse fiber length distribution. These structural features explain why the range of stiffnesses is larger for the experimental and random geometric graph (RGG) networks than for lattice-based networks. The geometrical disorder of the experimental and RGG networks also introduces an anomalous fold-stiffening, i.e. a dependence of stiffness on probe length that exceeds that predicted by continuum elasticity.

Our ``ideal-mechanosensor'' model does not address the internal mechanics of cells. Any internal noise in sensing or downstream signaling can only increase overall measurement uncertainty. There is evidence that cells can fix and thus regulate the relative uncertainty in measured stiffness by linearly modulating their applied stress to maintain a constant deformation (8). Notwithstanding, we have shown that even if cells sense nearly optimally, any single stiffness measurement is poorly informative of global tissue properties. This suggests that cells can benefit more from integrating the results of multiple stiffness measurements than from optimizing individual measurements, which may explain why some cells display more than a hundred focal adhesions (33). Yet even this strategy has diminishing returns, because nearby stiffness measurements probe the same underlying local structure and are therefore correlated. To extract useful information, cells must spread their measurements over extended regions of space, either by moving or by extending their shape. The benefits of an elongated shape are twofold, since measurements on larger scales are both more accurate and less spatially correlated. The biological relevance of this strategy is supported by the observation of highly polarized cells over five times longer than wide, including fibroblasts, mesenchymal stem cells, and cancer cells (5, 13, 33, 34).

Since cells and the ECM evolved together, the latter may also be tuned to facilitate mechanical inference. We uncovered two design principles for tuning the mechanical response: proximity to an elastic transition and the presence of long fibers. Both features result in an enhanced fold-stiffening for shorter probes, which may be harnessed to provide specialized signals to different cell types that measure over different scales, albeit the fold-stiffening comes at the cost of increased variance of the local stiffness. With regard to this variance, while we have only considered isotropic networks, actual tissues can have aligned fibers (12), which may yield a narrower stiffness distribution when probed along a fixed direction. This may explain why the alignment of the ECM has been observed to coordinate with cell polarization to promote migration (35). \\
\indent Finally, probing stiffness beyond linear response could provide cells with additional information. How hard would a cell need to pull to access the nonlinear regime? Nonlinear effects will certainly become significant when fibers begin to buckle (36). For a single fiber equal in length to the mesh size, the Euler buckling thresholds are $0.4$ and $2$ nN using the bending moduli we inferred for the collagen and fibrin networks, respectively. Forces of several newtons are achievable by stronger cells, as well as by colonies that migrate collectively (37). This suggests that the nonlinear regime is an interesting direction for future study. \\
\indent In summary, the disorder inherent in biological fiber networks places severe physical limits on the accuracy of cellular mechanosensing, suggesting that organisms must have evolved cellular-scale strategies to cope with this uncertainty \emph{in vivo}. Going forward, high-throughput gene deletion and mutation studies in conjunction with realistic patterned substrates (38) can help reveal the full array of internal components and pathways required for accurate mechanosensation.

\section{Network-based model for mechanosensing}
\label{sec:model} 
Fibers are modeled as simple elastic elements with a stretching modulus $\mu$
  (units of energy/length) and bending modulus $\kappa$
  (units of energy$\times$
 length). Adding the energy contributions from the stretching and bending of all fibers gives the Hamiltonian for the total mechanical energy of the network to leading order for small deformations:
\begin{multline}
H_{\mathrm{lin}}=\frac{\mu}{2}\sum\limits_{\langle ij\rangle}\frac{\left(\boldsymbol{u}_{ij}\cdot\boldsymbol{\hat{r}}_{ij}\right)^{2}}{\ell_{ij}} \\
+\frac{\kappa}{2}\sum\limits_{\langle ijk\rangle}\frac{1}{\ell_{ijk}}\left(\frac{\boldsymbol{u}_{jk}\times\boldsymbol{\hat{r}}_{ij}}{\ell_{jk}}-\frac{\boldsymbol{u}_{ij}\times\boldsymbol{\hat{r}}_{jk}}{\ell_{ij}}\right)^{2},
\end{multline}
\noindent where $\ell_{ij}$ is the length of the fiber $ij$ connecting
vertices $i$ and $j$ in the unperturbed reference state, $\ell_{ijk}=\left(\ell_{ij}+\ell_{jk}\right)/2$
is the average unstretched length of fibers $ij$ and $jk$, $\boldsymbol{u}_{ij}=\boldsymbol{u}_{j}-\boldsymbol{u}_{i}$
is the difference in vertex displacements $\boldsymbol{u}_{i}$, and
$\boldsymbol{\hat{r}}_{ij}$ is the unit vector in the direction of
the fiber in the undeformed reference state. Here, the first sum corresponds
to stretching interactions and is taken over all fibers $ij$. The
second sum corresponds to bending interactions and is taken over all
connected pairs of fibers $ij$ and $jk$. \newline
\indent The cell is modeled as an idealized stiffness-measuring device that
exerts a contractile, equal and opposite force of magnitude $f_0$
on vertices $1$ and $2$, which are separated by a distance $d$ in
the undeformed network. The vertices are chosen such that they are not directly connected to each other, in order to avoid trivial contributions to the measured stiffness due to the stretching characteristics of a single fiber. Provided the vertices $1$ and $2$ belong
to a rigid structure, in linear response the applied force deforms the network according
to the force-balance equations $\mathcal{D}_{ij}\boldsymbol{u}_{j}=\boldsymbol{f}_{i}$, where $\mathcal{D}_{ij}$ is the force-constant matrix of
the unperturbed Hamiltonian. The measured
local stiffness $k_{\mathrm{loc}}$ is defined as the magnitude of
the applied force divided by the relative displacement along the direction
of the force $k_{\textrm{loc}}\equiv - f_0 / (\boldsymbol{u}_{12}\cdot\boldsymbol{\hat{r}}_{12}$). We obtain the local stiffnesses numerically by computing the deformations
of the vertices $\boldsymbol{u}_{i}$ using the generalized inverse
of the force-constant matrix (\emph{Supporting Information}).

\begin{acknowledgments}
We thank Yigal Meir for insightful discussions. This work was supported in part by the National Science Foundation Grants DMR-1310266 (to L.M.J., S.M., and D.W), the Harvard Materials Research and Engineering Center DMR-1420570 (to L.M.J., S.M., and D.W.), a Lewis-Sigler fellowship (to C.P.B.), the German Excellence Initiative via the program ``NanoSystems Initiative M\"unich'' (NIM) and the Deutsche Forschungsgemeinschaft (DFG) via project B12 within the SFB-1032 (to C.P.B. and F.B.), National Science Foundation Grants PHY-1305525 and PHY-1066293 (to C.P.B., F.B., and N.S.W.), and the hospitality of the Aspen Center for Physics (to C.P.B. and N.S.W.).
\end{acknowledgments}

\newpage

\renewcommand{\theequation}{S\arabic{equation}}

\titleformat{\section}    
       {\normalfont\fontfamily{cmr}\fontsize{12}{17}\bfseries}{\thesection}{1em}{}
\titleformat{\subsection}[runin]
{\normalfont\fontfamily{cmr}\bfseries}{}{1em}{}

\renewcommand\thefigure{S\arabic{figure}}    
\renewcommand\thesection{S\arabic{section}}

\onecolumngrid

\setcounter{section}{0}
\setcounter{figure}{0}
\setcounter{equation}{0}

\part*{\centerline{Supporting Information}}

\section{\textbf{Mechanosensing model}}

\subsection{Fiber network model.}

We study the mechanical properties of crosslinked biopolymer networks, such as those of the extracellular matrix (ECM),
using a fiber network model, which consists of a collection of elastic
elements that are connected at point-like vertices. The elastic elements
model the stretching and bending interactions of the constituent fibers.
The mechanical energy in the fiber network model is given by (29):

\begin{equation}
H=\frac{\mu}{2}\sum_{\langle ij\rangle}\frac{\left(|\boldsymbol{x}_{ij}|-\ell_{ij}\right)^{2}}{\ell_{ij}}+\frac{\kappa}{2}\sum_{\langle ijk\rangle}\frac{\left(\theta_{ijk}-\theta_{ijk}^{(0)}\right)^{2}}{\ell_{ijk}},
\end{equation}

\noindent where

\begin{itemize}

\item $\mu$ is the stretching modulus,
\item $\kappa$ is the bending modulus,
\item $\boldsymbol{x}_{ij}=\boldsymbol{x}_{j}-\boldsymbol{x}_{i}$ is the
difference in positions of vertices $i$ and $j$,
\item $\ell_{ij}$ is the length of the fiber $ij$
connecting vertices $i$ and $j$ in the unperturbed reference state (i.e. the unstretched fiber length),
\item $\ell_{ijk}=\left(\ell_{ij}+\ell_{jk}\right)/2$ is the average unstretched
length of fibers $ij$ and $jk$,
\item $\theta_{ijk}=\sin^{-1}\left(|\hat{\boldsymbol{x}}_{ij}\times\hat{\boldsymbol{x}}_{jk}\right|)$
is the angle between fibers $ij$ and $jk$
that meet at vertex $j$,
\item $\theta_{ijk}^{(0)}=\sin^{-1}\left(|\hat{\boldsymbol{r}}_{ij}\times\hat{\boldsymbol{r}}_{jk}\right|)$
is the angle between unit vectors $\hat{\boldsymbol{r}}_{ij}$ and
$\hat{\boldsymbol{r}}_{jk}$ that point along the direction of the
fibers connecting vertices in the unperturbed reference state.

\end{itemize}

\noindent The first sum corresponds to stretching interactions and is taken
over all fibers $ij$. The second sum corresponds to bending interactions
and is taken over all connected pairs of fibers $ij$ and $jk$.

Small forces will generate deformations that are small compared to
the lengths of fibers. In this case, it is convenient to express the
positions of the vertices in terms of their deformations $\boldsymbol{u}_{i}$
about their positions in the undeformed reference state $\boldsymbol{r}_{i}$
as follows:

\begin{equation}
\boldsymbol{x}_{i}=\boldsymbol{r}_{i}+\boldsymbol{u}_{i}.
\end{equation}

\noindent Upon expanding both sums in the mechanical energy and discarding higher
order terms, we obtain the harmonic energy in the fiber network
model, which is given by:

\begin{equation}
H_{\mathrm{lin}}=\frac{\mu}{2}\sum_{\langle ij\rangle}\frac{1}{\ell_{ij}}\left(\boldsymbol{u}_{ij}\cdot\hat{\boldsymbol{r}}_{ij}\right)^{2}+\frac{\kappa}{2}\sum_{\langle ijk\rangle}\frac{1}{\ell_{ijk}}\left(\frac{\boldsymbol{u}_{jk}\times\hat{\boldsymbol{r}}_{ij}}{\ell_{jk}}-\frac{\boldsymbol{u}_{ij}\times\hat{\boldsymbol{r}}_{jk}}{\ell_{ij}}\right)^{2},
\end{equation}

\noindent where $\boldsymbol{u}_{ij} = \boldsymbol{u}_{j} - \boldsymbol{u}_{i}$ is the relative deformation of nodes $i$ and $j$.

\subsection{Idealized measurement device.}

The cell is modeled as an idealized stiffness-measuring device that
exerts force on vertices of the network. The effect of the applied
force $\boldsymbol{f}_{i}$ perturbs the mechanical energy as follows:

\begin{equation}
\delta H= - \sum_{i}\boldsymbol{f}_{i}\cdot\boldsymbol{u}_{i},
\end{equation}

\noindent where the index is summed over all vertices. We assume that the ECM behaves as a viscoelastic solid, and that the forces applied by cells change slowly. Assuming
forces applied by cells change slower than the timescale set
by viscous damping, the network will be deformed in a quasistatic
manner, i.e. the network reaches mechanical equilibrium for a given
force. In this case, the deformation is completely determined by minimizing
the full mechanical energy:

\begin{equation}
\frac{d}{d\boldsymbol{u}_{i}}\left(H+\delta H\right)=0.
\end{equation}

\noindent Expressing this equation in terms of the deformations and
the forces leads to the equations of force-balance in static equilibrium:

\begin{equation}
\sum_{j}\mathcal{D}_{ij}\boldsymbol{u}_{j}=\boldsymbol{f}_{i},
\end{equation}

\noindent where $\mathcal{D}_{ij}$ is the force-constant matrix of
the unperturbed Hamiltonian:

\begin{equation}
\mathcal{D}_{ij}\equiv\frac{\partial H}{\partial\boldsymbol{u}_{i}\partial\boldsymbol{u}_{j}}.
\end{equation}

We model a two-point measurement performed by a cell as a force dipole,
which is defined as a contractile force exerted on vertices $1$ and
$2$ separated by a vector $\boldsymbol{d}=d\ \hat{\boldsymbol{r}}_{12}$
in the undeformed reference state. The vector of equal and opposite
forces may be represented using Kronecker delta notation as follows:

\begin{equation}
\boldsymbol{f}_{i}=f_{0}\left(\delta_{i1}-\delta_{i2}\right) \hat{\boldsymbol{r}}_{12},
\end{equation}

\noindent where $f_{0}$ is the magnitude of the force applied to
each vertex.

The local stiffness $k_{\mathrm{loc}}$ is defined as the linear response
of the displacement of vertices $1$ and $2$ to the contractile force:

\begin{equation}
k_{\textrm{loc}}\equiv\frac{-f_{0}}{\boldsymbol{u}_{12}\cdot \hat{\boldsymbol{r}}_{12}}.
\end{equation}

\subsection{Numerical procedure.}

To calculate the local stiffness $k_{\mathrm{loc}}$ defined above,
we compute the deformations of the vertices $\boldsymbol{u}_{i}$
numerically by solving the equations of force-balance in static equilibrium.
These equations do not have a well-defined solution for all possible
configurations of the applied force because the force-constant matrix
is singular, i.e. it contains eigenvectors with vanishing eigenvalues,
or ``zero modes.'' If the force of a local stiffness probe couples
to a zero mode, the resulting deformations will diverge and the local
stiffness is undefined in linear response. Intuitively, this corresponds to probes that act on unconstrained, dangling portions of the network.

To solve the equations of force-balance while dealing with the technical
challenge provided by zero modes, we compute the generalized inverse
of the force-constant matrix. The generalized inverse allows us to
first check whether a solution exists for a given force perturbation,
and then to solve for the deformations by multiplying the force perturbation
by the generalized inverse. Such an approach is efficient because
it allows us to measure local stiffness over an entire network using
only a single matrix multiplication operation per probe, which is
computationally inexpensive compared with solving the system of linear
equations anew for each local stiffness probe.

We obtain the generalized inverse by first computing the singular
value decomposition (SVD) of the force-constant matrix:

\begin{equation}
\mathcal{D}\Rightarrow\mathcal{USV}^{T},
\end{equation}

\noindent where $\mathcal{U}$ and $\mathcal{V}$ are orthogonal matrices
and $\mathcal{S}$ is a diagonal matrix consisting of singular values
$\left\{ s_{1},\ s_{2},\ldots,\ s_{r}\right\} $ equal in number to
the rank $r$ of the dynamical matrix. From this, the generalized
inverse may be computed:

\begin{equation}
\mathcal{D}^{+}=\mathcal{VS}^{+}\mathcal{U}^{T},
\end{equation}

\noindent where $\mathcal{S}^{+}$ is a diagonal matrix consisting
of the reciprocals of the singular values $\left\{ \frac{1}{s_{1}},\ \frac{1}{s_{2}},\ldots,\ \frac{1}{s_{r}}\right\} $.
The generalized inverse provides a simple way to check whether a solution
exists. That is, if a given force vector satisfies the following equation:

\begin{equation}
\mathcal{D}\mathcal{D}^{+}\boldsymbol{f}=\boldsymbol{f},
\end{equation}

\noindent then the force-balance equation has a solution. This guarantees that a given force configuration has no projection
onto any zero mode. For well-defined force probes, which satisfy (S12), the resulting deformation
of each vertex is finite and given by:

\begin{equation}
\boldsymbol{u}_{i}=\sum_{j}\mathcal{D}_{ij}^{+}\boldsymbol{f}_{j},
\end{equation}

\noindent from which the local stiffness may be computed.

\newpage

\section{Analysis of network architectures}

\vspace{10 pt}

\subsection{Experimental network architectures.}

%FigS1
\begin{figure*}[t!]
\centerline{
\includegraphics[width=\columnwidth]{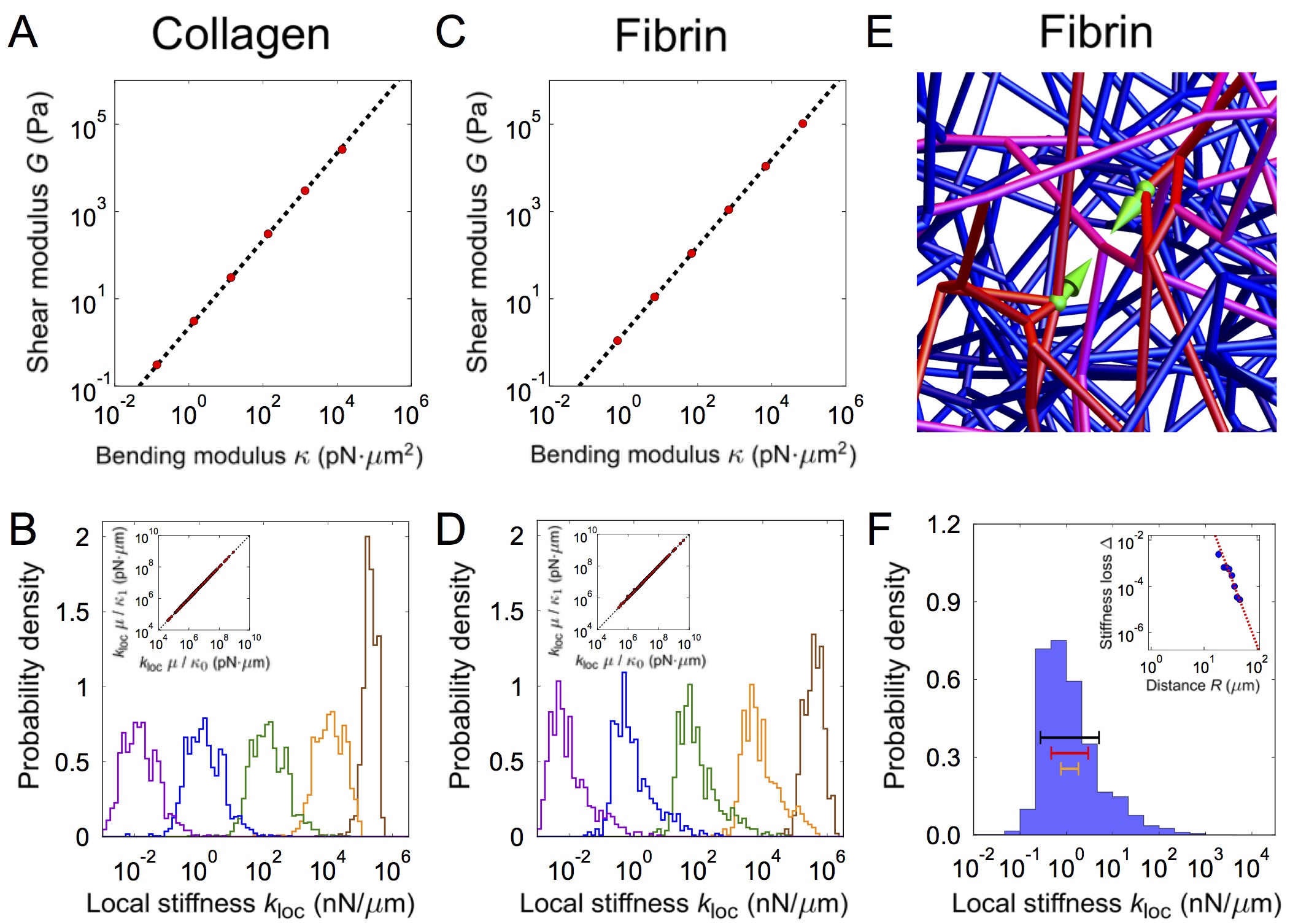}
}
\caption{\label{fig:F1}
Analysis of experimental network architectures. (\emph{A} and \emph{C}) Macroscopic shear modulus $G$ versus bending modulus $\kappa$ for (\emph{A}) experimental collagen network and  (\emph{C}) experimental fibrin network. (\emph{B} and \emph{D}) Distribution of local stiffnesses $k_{\mathrm{loc}}$ for a broad range of bending moduli for (\emph{B}) experimental collagen network ($\kappa$: $\kappa_0=14$ nN$ \cdot \mu$m$^2$ purple, $10^{2}\kappa_0$ blue, $10^{4}\kappa_0$ green, $10^{6}\kappa_0$ orange, and $10^{8}\kappa_0$ brown) and (\emph{D}) experimental fibrin network ($\kappa$: $\kappa_0=75$ nN$ \cdot \mu$m$^2$ purple, $10^{2}\kappa_0$ blue, $10^{4}\kappa_0$ green, $10^{6}\kappa_0$ orange, and $10^{8}\kappa_0$ brown). \emph{Insets}: comparison of stiffness measurements for two different values of bending modulus, $\kappa_0$ and $10^{2}\kappa_0$, scaled by bending modulus for (\emph{B}) collagen network ($\kappa_0=14$ nN$ \cdot \mu$m$^2$) and (\emph{D}) fibrin network ($\kappa_0=75$ nN$ \cdot \mu$m$^2$). Dashed line shows linear fit. (\emph{E}) Modeled deformation under stress from a local force dipole of length $d \sim 15 \mu$m (green arrows) of experimental fibrin network. Magnitude of fiber deformations indicated by color (small deformations blue, large deformations red). (\emph{F}) Distribution of local stiffnesses $k_{\mathrm{loc}}$ (force dipole of length $d \sim 15 \mu$m) for experimental fibrin network. Geometric standard deviation of local stiffness $\sigma_{\mathrm{loc}}$ indicated by bars (actual distribution black, estimated distribution assuming strong locality red, estimated distribution assuming weak locality orange. \emph{Inset}: stiffness loss ${\Delta}$, defined as the relative change in local stiffness $k_{\mathrm{loc}}$ upon perturbing a network by removing a single fiber, versus distance $R$ of center of removed fiber from probe center (probe length $d<10 \mu$m and removed-fiber length $\ell_{ij}<10\mu$m). Error bars are smaller than the size of data points. Dashed lines show asymptotic scaling from continuum theory, which predicts ${\Delta} \sim 1/R^{2D}$ for $R >> d$.}
\end{figure*}

The network architectures for the fiber network models, i.e. the positions
of the vertices and their pairwise connectivities, were obtained from
experiment by analyzing images of reconstituted fiber networks. Collagen networks were prepared and imaged as described previously in (36). Briefly, a sample of 0.2 mg/ml protein in 1x DMEM with 25mM HEPES was prepared from a solution of collagen type-I monomers with a small fraction of fluorescently-labelled monomers at 4$^{\circ}$ C. Network formation was induced by neutralizing the sample's pH with 1M NaOH and incubating it at room temperature for four hours. The resulting network was imaged using a confocal microscope (Leica SP5, Wetzlar, Germany). We acquired a set of fluorescent images covering a three-dimensional volume that was representative of the network structure. To determine the fiber positions and connectivity, the image stacks were thresholded and subsequently skeletonized, which resulted in a one-voxel thick line representation of all fibers. We define a branch-point as the junction between three or more fibers. The number of fibers that join at a branch-point defines the branch-point's connectivity $z$. In our mechanical model, the vertices are positioned at the branch-points and end-points of fibers. By counting the number of vertices within the network volume, we found a vertex density of $8.52\cdot10^{-4}$ vertices/$\mu\mathrm{m}^{3}$.

Fibrin networks were prepared and imaged as described previously (44). Briefly, a solution of human fibrinogen (Enzyme Research Labs, South Bend, IN) containing a small fraction of fluorescently labelled fibrinogen with a protein concentration of 0.2 mg/ml was prepared in buffer (150mM NaCl, 20mM CaCl, 20mM HEPES, pH 7.4). Network formation was induced by the addition of activated human alpha-thrombin (Enzyme Research Labs, South Bend, IN) to the fibrinogen solution (final thrombin concentration: 0.1 U/ml). After the sample was allowed to polymerize for 12 hours, the resulting fibrin network was subsequently imaged and the data was processed analogously to the collagen sample. The vertex density was around $8.07\cdot10^{-4}$ vertices/$\mu\mathrm{m}^{3}$.

For the collagen and fibrin networks, we found average coordination numbers of $z\simeq2.9$ and $z\simeq2.7$, respectively. Although these average coordination numbers are low, the networks are both macroscopically rigid, which is defined by a finite response to shear forces. Rheological measurements were performed on networks prepared under experimental conditions analogous to those of the imaged samples. For the collagen network, the measurements were done using an AR-G2 rheometer (TA instruments, New Castle DE) equipped with a custom-made plastic-plate of 25mm and sealed with mineral oil (44). The macroscopic mechanical response was measured by applying a small oscillatory strain and measuring the resulting stress, which resulted in a shear modulus of $G\simeq0.3$ Pa. The fibrin network was measured with a 40 mm / $4^{\circ}$ cone-plate geometry. Samples were sealed with mineral oil and allowed to polymerize for at least 12 hours. The network was then perturbed in the same manner as for the collagen network and the shear modulus was found to be $G\simeq1$ Pa.

Macroscopic rigidity requires macroscopic connectedness,
i.e. the presence of a spanning cluster of vertices. In our analysis,
we identified all vertices that belong to the largest cluster using
a union-find algorithm and removed all other vertices from the network. For both the collagen and the fibrin networks, we considered a spherical
sample of radius $R_{\textrm{sample}}=65\ \mu$m. Within this sample volume, the $N$ vertices
are distributed approximately homogeneously on scales large compared
to the mesh size $\xi$, defined as the radius of a sphere
whose volume is equal to the average volume per vertex:

\begin{equation}
\xi=\frac{R_{\textrm{sample}}}{N^{1/3}}.
\end{equation}

\noindent For the experimental collagen and fibrin networks, we found $\xi \simeq 6.5\ \mu$m and $6.7\ \mu$m, respectively. On scales comparable to $\xi$, the networks are intrinsically disordered due to heterogeneity in both the spatial positions of the vertices and the lengths of fibers.
The lengths of the fibers are highly polydisperse, spanning over an
order of magnitude in length with a substantial fraction of fibers that
are very long compared to the average fiber length (Fig. S9\emph{A},\emph{B}). The fiber length distribution peaks at small lengths (i.e.
below the average fiber length) and decays roughly monotonically beyond
the peak.

For each experimental network, the corresponding mechanical model has two undetermined parameters: the stretching modulus $\mu$ and the bending modulus $\kappa$. Since both parameters may be rescaled together resulting only in a trivial change of units, the only remaining non-trivial parameter is their ratio $\kappa/\mu$. Throughout the main text, we chose $\kappa/\mu = 10^{-5}$ to capture the fact that individual fibers are much stiffer to stretching than to bending.

We compute the local stiffness distribution numerically using the
procedure described above in Section 1.3; however, rather than considering
an ensemble of individual probes over many network realizations, we
sample many probes throughout the volume of a single network. These
procedures are equivalent when the sample volumes are large compared
to the scale over which local stiffness measurements are correlated,
i.e. networks which are large enough to represent the ensemble via
self-averaging. We confirmed this equivalence by varying the sample volume and
observing that the distribution converged below the network radius
used in our analysis.

To determine the units of the modeled response, we first noted that both the experimental collagen and fibrin networks display a bending-dominated response. That is, for these networks, the macroscopic shear modulus and all local stiffness measurements are directly proportional to $\kappa$ (Fig. S1\emph{A}-\emph{D}). We therefore fit the bending modulus of our mechanical model using the experimentally-measured macroscopic response. Specifically, we first calculated the shear modulus numerically using the mechanical model for an arbitrary value of $\kappa$, which we then scaled by the ratio of the experimental shear modulus to the calculated shear modulus. This resulted in fitted bending moduli of $\kappa=14$ nN$ \cdot \mu$m$^2$ and $\kappa=75$ nN$ \cdot \mu$m$^2$ for collagen and fibrin, respectively.

\subsection{Modeled network architectures.}

To understand how the mechanical response emerges from the features
of the network architecture, we considered idealized model architectures. For simplicity, the resulting structures were generated randomly such that the initial positions of the vertices and fibers were not spatially correlated. Throughout our analysis, we obtained our results for these networks by averaging over a large number of network realizations to reduce sampling noise.

\subsubsection{Random geometric graph model}

Random geometric graphs (RGG) provide a way to generate model architectures
with independent control over average coordination number, vertex
density, and average fiber length. In particular, these parameters
can be chosen to match those observed in experiment. In the RGG model,
architectures are generated by first distributing vertices throughout
a volume and then connecting them randomly according to a specified probability
distribution. The position of each vertex is drawn
from a uniform distribution over the network volume, and the probability distribution for connecting
two vertices, or the ``connection function,'' depends only on the inter-vertex distance. For simplicity, we model the connection function as a power law with
an exponential cutoff:

\begin{equation}
P_{c}(\ell)\propto\frac{e^{-\ell/L}}{\ell^{2}},
\end{equation}

\noindent where $L$ is a length scale that describes the typical
length of fibers. This results in networks that are homogeneous and isotropic over large scales. For such networks in three dimensions, the expected density of vertices that are a distance $\ell$ from a
particular vertex scales as $\ell^{2}$, so the expected distribution
of fiber lengths in the three-dimensional RGG model is $P(\ell)\propto\ell^{2}P_{c}(\ell)\propto e^{-\ell/L}$.

Throughout our analysis, we considered networks with approximately
$6.25\cdot10^{-3}$ vertices/$\mu\mathrm{m}^{3}$ and we chose the
length scale $L$ equal to $3.5 \ \mu$m. We chose the average coordination number
$z\simeq2.9$ to match that of the experimental networks (except in
Fig. 2\emph{A} where we varied the average coordination number by randomly removing fibers from a network with a high initial average coordination number). This resulted in an average mesh size $\xi\simeq6.7\ \mu$m. Due
to the larger resulting density of these modeled networks in comparison
to the experimental networks, we considered samples of radius $R_{\textrm{sample}}=33\ \mu$m. We determined the bending modulus $\kappa$
by fitting the geometric mean stiffness measured for the RGG network at $d=15 \ \mu$m
to that of the collagen network. This resulted in a bending modulus
$\kappa=83$ nN$\ensuremath{ \cdot \mu}$m$^2$ and shear modulus $G=9\ \textrm{Pa}$. For the above choices of parameters, the RGG networks quantitatively capture the local mechanical response of the experimental networks, including the broad width of the local stiffness distribution, the large fold-stiffening for short probes, and the two-point stiffness correlation length (see \emph{Results}). Because of the broad range of stiffnesses, throughout our work, we characterize the variability of the stiffness distribution using the geometric standard deviation, defined as the exponential of the standard deviation of log-local stiffnesses:

\begin{equation}
\sigma_{\mathrm{loc}} = e^{\sqrt{\langle (\log k_{\mathrm{loc}} - \langle \log k_{\mathrm{loc}} \rangle)^2 \rangle}}
\end{equation}

\subsubsection{Disordered lattice model}

Disordered lattice networks are an established class of modeled architectures that have successfully described many of the mechanical properties of biopolymer networks (20, 22, 23). In particular, lattice models are able to capture the macroscopic response and its striking dependence on network density. Because of their numerical convenience, two-dimensional lattice-based
networks have provided a starting point for modeling the microscopic response of fiber networks (27). Here, we provide the
first numerical analysis of the microscopic response of three-dimensional
disordered lattice networks and demonstrate that such models fail to quantitatively capture the broad width we observed for the experimental collagen and fibrin networks.

Disordered lattice networks are generated by placing vertices on an ordered lattice, for example face-centered-cubic (FCC) or simple-cubic (SC) lattices in three dimensions, and
connecting them randomly with a fixed probability $p$. For a given choice of lattice, the connectivity $p$ is the sole parameter that impacts the overall network structure. In particular, the density of vertices cannot be varied independently from the average fiber length. Furthermore, since the vertices are initially positioned to lie on a regular lattice, fibers only meet at discrete, fixed angles in the unperturbed reference state. Disordered lattice networks are therefore significantly simpler than the experimental networks.

The local stiffness distribution for the FCC lattice network is shown in Fig. S2\emph{B}. Here, we chose a bond-probability of $p=0.3$
to study the bending regime relevant to biological networks. Throughout
our analysis, we set the stretching modulus $\mu=1$ and bond length $\ell_{0}=1$. We considered local
stiffness measurements for pairs of vertices separated by primitive lattice vectors of the form $(-1,\ 2,\ n>0)$
to avoid pathological effects associated with
applying forces in the symmetry directions of the lattice.

For both the SC and FCC disordered lattice networks, we find the presence of a broad range of local stiffnesses spanning up to roughly two decades (Fig. S4\emph{F}, Fig. S2\emph{B}). This broad range of stiffnesses is generic over a wide range of network connectivities in the stretching and bending regimes and peaked near the crossovers (see Fig. S2\emph{C} for FCC lattice results). Interestingly, although the range is qualitatively very large, the geometric standard deviation is substantially smaller for both the SC ($\sigma_{\mathrm{loc}}=0.29$, Fig. S4\emph{F}) and FCC ($\sigma_{\mathrm{loc}}=0.37$, Fig. S2\emph{B}) lattice networks than for the experimental (collagen: $\sigma_{\mathrm{loc}}=0.54$, Fig. 1\emph{C}; fibrin: $\sigma_{\mathrm{loc}}=0.63$, Fig. S1\emph{F}) and random geometric graph (RGG, $\sigma_{\mathrm{loc}}=0.56$, Fig. 1\emph{D}) networks. In the following section, we study how this discrepancy arises from differences in network structure.

What is the physical origin of the qualitatively broad width we observed for the disordered lattice networks? As for the RGG network (Fig. S5\emph{D}-\emph{F}), the disordered lattice networks also yield macroscopic shear moduli and median stiffnesses that depend sensitively on average coordination number (Fig. S4\emph{D}-\emph{F}). This suggests that the broad width we observed for all types of networks we studied arises from intrinsic features common to all types of disordered networks. Indeed, as we observed for the RGG network (see \emph{Results}), we found that for the disordered lattice networks: (1) the stiffness loss also obeys universal scaling (see Fig. S2\emph{B}, \emph{Inset} for FCC lattice results), and (2) the scaling of the marginal distribution of local stiffnesses with the number of local fibers $N_F$ is captured by the rapid, nonlinear scaling of the macroscopic shear modulus (Fig. S4\emph{D}-\emph{F}, \emph{Insets}).

We then estimated the geometric standard deviation for the SC and FCC networks in the same way as for the experimental and RGG networks (see next section, \emph{Dependence of local stiffness on local structure}). As before, we estimate the local stiffness distribution as the distribution of the number of local fibers $N_F$ transformed by the macroscopic shear modulus. For both SC and FCC lattice networks, we found that counting the number of local fibers using rapidly-decaying weighting functions results in a good match between the estimated and actual geometric standard deviations (Table S1, Fig. S4\emph{D}-\emph{F}). Furthermore, we found that the estimate does not depend significantly on the particular form of the weighting function used to calculate the number of local fibers. Specifically, the estimate of the geometric standard deviation for the FCC network does not depend strongly on whether we chose a rapidly decaying power-law or a hard cutoff at two mesh sizes (Fig. S4\emph{D},\emph{E}). This suggests that obtaining a good estimate of the geometric standard deviation based on local structure does not require finely-tuning the details of the weighting function, provided the chosen weights are given by a function that decays sufficiently rapidly with distance.

%FigS2
\begin{figure*}[t!]
\centerline{
\includegraphics[width=\columnwidth]{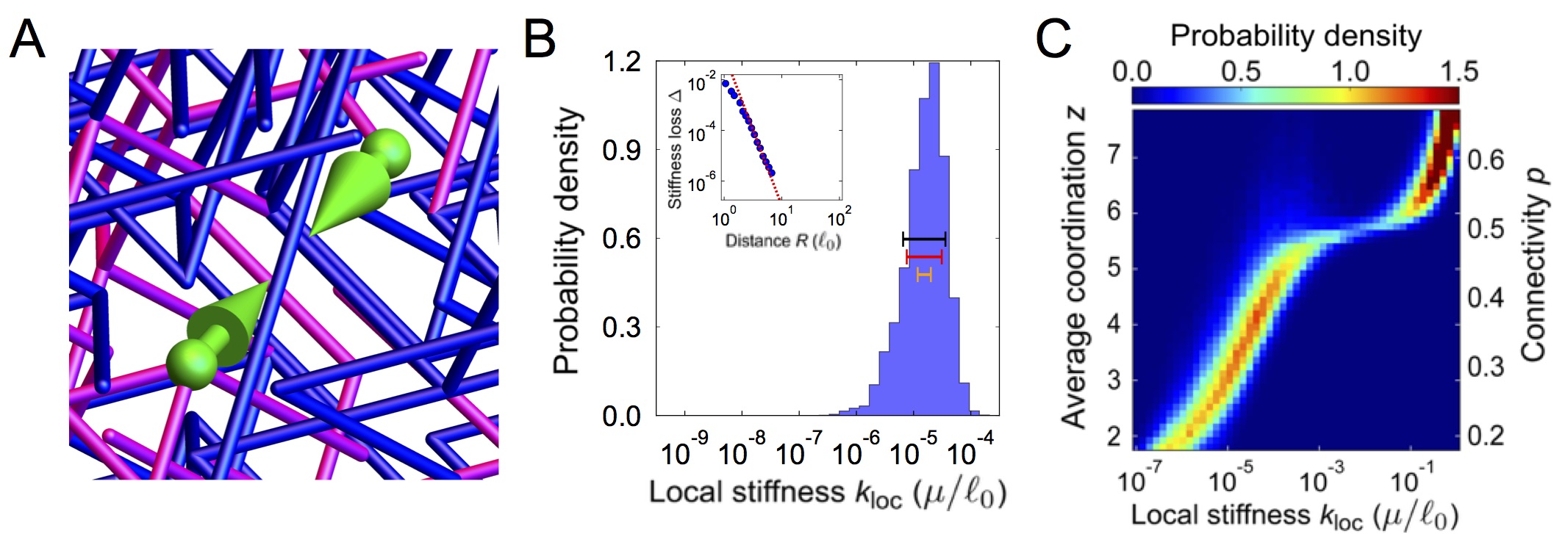}
}
\caption{\label{fig:F2}
Analysis of FCC lattice network. (\emph{A}) Modeled deformation under stress from a local force dipole (green arrows) of disordered face-centered-cubic lattice network (FCC, $d = \sqrt{5} \ell_0$, bond occupancy $p=0.3$, and bending modulus $\kappa=10^{-7} \ \mu \ell_0^2$). Magnitude of fiber deformations indicated by color (small deformations blue, large deformations red). (\emph{B}) Distribution of local stiffnesses $k_{\mathrm{loc}}$ for FCC network. Geometric standard deviation of local stiffness $\sigma_{\mathrm{loc}}$ indicated by bars (actual distribution black, estimated distribution assuming strong locality red, estimated distribution assuming weak locality orange). \emph{Inset}: stiffness loss $\Delta$, defined as the relative change in local stiffness $k_{\mathrm{loc}}$ upon perturbing a network by removing a single fiber, versus distance $R$ of center of removed fiber from probe center ($d = \sqrt{5} \ell_0$). Error bars are smaller than the size of data points. Dashed lines show asymptotic scaling from continuum theory, which predicts $\Delta \sim 1/R^{2D}$ for $R >> d$. (\emph{C}) Distribution of local stiffnesses $k_{\mathrm{loc}}$ for FCC network versus average coordination number of vertices, $z$, or equivalently, bond occupancy $p$ (force dipole length $d=\sqrt {5} \ell_0$, bending modulus $\kappa = 10^{-7} \ \mu \ell_0^2$.
}
\end{figure*}

\clearpage

\section{Dependence of local stiffness on local structure}

%FigS7
\begin{figure*}[t!]
\centerline{
\includegraphics[scale=0.15]{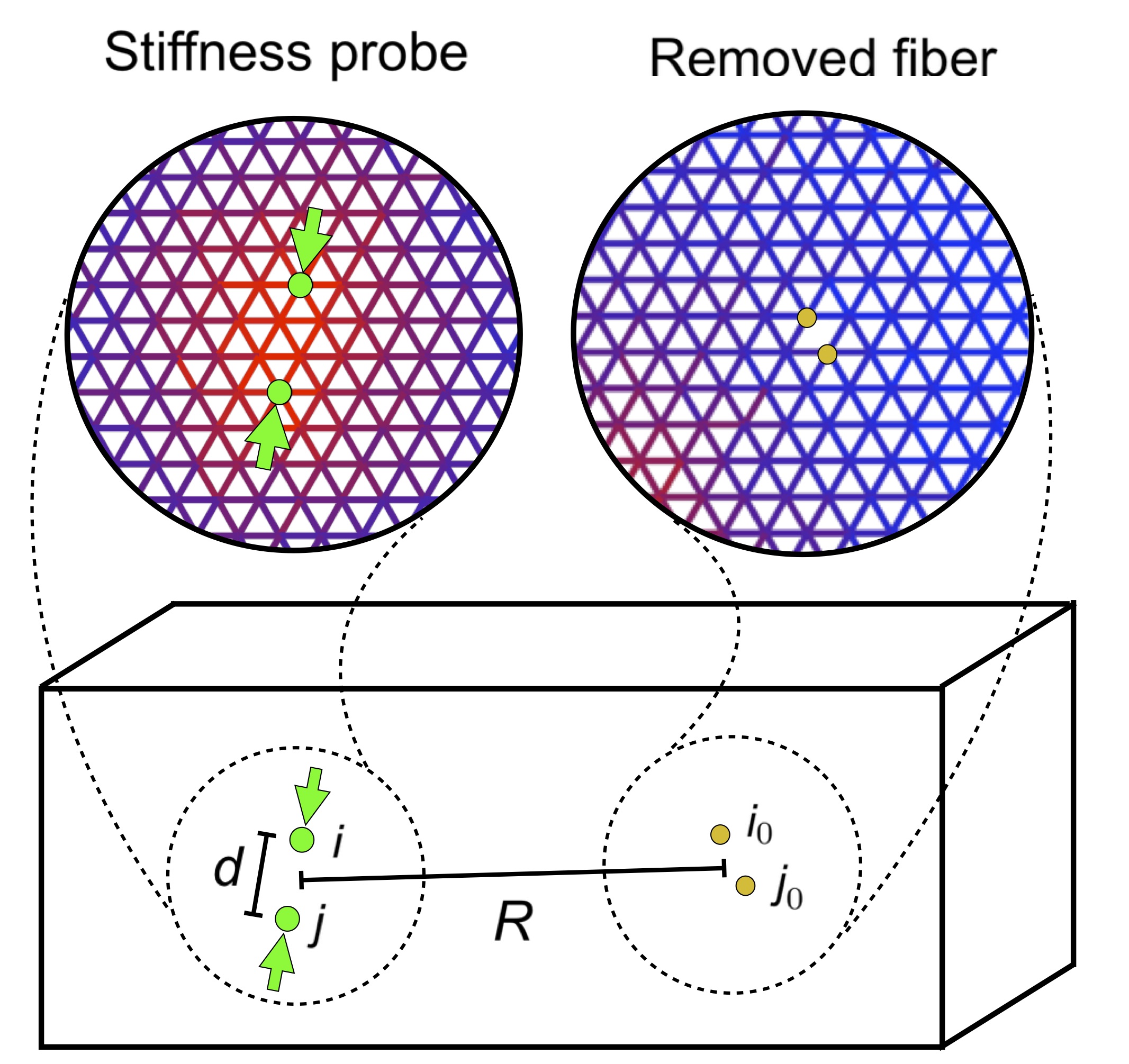}
}
\caption{\label{fig:F7}
Stiffness loss due to fiber removal. Schematic illustration of the geometry used to calculate the fiber-removal effect. The fiber-removal effect is the relative change in local stiffness when a single fiber is removed at a distance $\textit{R} \gg d$.}
\end{figure*}

To gain insight into the variation of local stiffness, we determined how the presence of a single fiber in the vicinity of
the stiffness probe impacts the local stiffness. We quantified the influence of a fiber by the stiffness
loss ${\Delta}$, defined as the relative change in local stiffness upon removing the
fiber from the network (Fig. S3). Intuitively, fibers which are more proximal should have a greater effect. We therefore expect the stiffness loss induced by removing a fiber to decay
as a function of the distance $R$ from the probe center to the center
of the fiber. In this section, we calculate how the stiffness loss
scales with distance $R$ for an ordered lattice.

\subsection{Stiffness loss scaling.}

We calculate the stiffness loss scaling by first deriving an exact
formula for the change in local stiffness when a single fiber is removed
from an ordered, stretching-only network in $D$ dimensions, and then studying how
the magnitude of the effect depends on the distance to the missing
fiber from the center of the probe. The equation of force-balance in static equilibrium (Eq. S6) becomes:

\begin{equation}
\mathcal{D}_{0}\cdot\boldsymbol{u}=\boldsymbol{f},
\end{equation}

\noindent where the unperturbed force-constant matrix $\mathcal{D}_{0}$,
vertex deformations, and applied forces are represented using Dirac
notation as follows:

\begin{equation}
\mathcal{D}_{0}=\mu\sum_{\langle ij\rangle}|i\rangle\left(b_{ij}\hat{\boldsymbol{r}}_{ij}\otimes\hat{\boldsymbol{r}}_{ij}-\delta_{ij}\sum_{j'\neq i}b_{ij'}\hat{\boldsymbol{r}}_{ij'}\otimes\hat{\boldsymbol{r}}_{ij'}\right)\langle j|
\end{equation}

\begin{equation}
\boldsymbol{u}=\sum_{i}|i\rangle\boldsymbol{u}_{i}
\end{equation}

\begin{equation}
\boldsymbol{f}=\sum_{i}|i\rangle\boldsymbol{f}_{i},
\end{equation}

\noindent where $b_{ij}=1$ if vertices $i$ and $j$ are nearest neighbors
and $b_{ij}=0$ otherwise, and the vectors $|i\rangle$ form a complete
orthonormal set over vertices such that $\langle i|k\rangle=\delta_{ik}$
and $\sum_{i}|i\rangle\langle i|=1$.

These equations can be inverted by Fourier transformation to solve
for the displacements:

\begin{equation}
\boldsymbol{u}=\mathcal{G}_{0}\cdot\boldsymbol{f},
\end{equation}

\noindent where $\mathcal{G}_{0}$ is the lattice Green's tensor
defined by $\mathcal{G}_{0}\mathcal{D}_{0}=1$ and given by

\begin{equation}
\mathcal{G}_{0}(i,j)\equiv\langle i|\mathcal{G}_{0}|j\rangle=\int_{k\in BZ}\frac{d^{D}k}{(2\pi)^{D}}e^{i\boldsymbol{k}\cdot\boldsymbol{r}_{ij}}\mathcal{D}_{0}^{-1}(\boldsymbol{k}),
\end{equation}

\noindent in terms of the $D\times D$ dynamical matrix: 
\begin{equation}
\mathcal{D}_{0}(\boldsymbol{k})=\mu\sum_{\hat{\delta}}\left(1-e^{i\ell_{0}\boldsymbol{k}\cdot\hat{\delta}}\right)\hat{\delta}\otimes\hat{\delta}.
\end{equation}

\noindent Here, $\hat{\delta}$ are unit vectors summed over the directions
of the nearest neighbor vertices. The lattice Green's tensor relates the force applied to vertex $i$ to the deformation
at vertex $j$, which provides a simple way to express the relative
deformation of the vertices induced by the dipole:

\begin{equation}
u_{0}=2f_{0}\hat{\boldsymbol{d}}\cdot\left[\mathcal{G}_{0}(i,i)-\mathcal{G}_{0}(i,j)\right]\cdot\hat{\boldsymbol{d}}.
\end{equation}

\noindent This relative deformation determines the local stiffness
defined in (Eq. S9) as the ratio of the applied force to the relative
deformation $k_{\mathrm{loc}}=f_{0}/u_{0}$.

We will now calculate the perturbed stiffness $\tilde{k}_{\mathrm{loc}}$
between vertices $i$ and $j$ when the bond $i_{0}j_{0}$ connecting
neighboring vertices $i_{0}$ and $j_{0}$ is removed, generalizing
the analogous calculation for resistor networks performed by Cserti et al. (40). The equation of force-balance for the perturbed network
is

\begin{equation}
\left(\mathcal{D}_{0}-\delta\mathcal{D}\right)\cdot\boldsymbol{u}=\mathcal{D}\cdot\boldsymbol{u}=\boldsymbol{f},
\end{equation}

\noindent where $\mathcal{D}=\mathcal{D}_{0}-\delta\mathcal{D}$ is
the force-constant matrix for the perturbed network, i.e. the unperturbed
force-constant matrix $\mathcal{D}_{0}$ minus the contributions associated
with the missing bond $i_{0}j_{0}$:

\begin{equation}
\delta\mathcal{D}=\boldsymbol{\hat{r}}_{i_{0}j_{0}}\otimes\boldsymbol{\hat{r}}_{i_{0}j_{0}}|i_{0}\rangle\langle j_{0}|-\boldsymbol{\hat{r}}_{i_{0}j_{0}}\otimes\boldsymbol{\hat{r}}_{i_{0}j_{0}}|i_{0}\rangle\langle i_{0}|+\boldsymbol{\hat{r}}_{j_{0}i_{0}}\otimes\boldsymbol{\hat{r}}_{j_{0}i_{0}}|j_{0}\rangle\langle j_{0}|-\boldsymbol{\hat{r}}_{j_{0}i_{0}}\otimes\boldsymbol{\hat{r}}_{j_{0}i_{0}}|j_{0}\rangle\langle i_{0}|
\end{equation}

\begin{equation}
=|x\rangle\boldsymbol{\hat{r}}_{i_{0}j_{0}}\otimes\boldsymbol{\hat{r}}_{i_{0}j_{0}}\langle y|,
\end{equation}

\noindent where $|x\rangle=|i_{0}\rangle-|j_{0}\rangle$ and $\langle y | =\langle i_{0}|-\langle j_{0}|$.
We solve for the deformations as before:

\begin{equation}
\boldsymbol{u}=\mathcal{G}\cdot\boldsymbol{f},
\end{equation}

\noindent in terms of $\mathcal{G}$, the lattice Green's tensor
for the perturbed network defined by $\mathcal{D}\cdot\mathcal{G}=1$.
By premultiplying this equation with $\mathcal{G}_{0}$, we obtain
a recursive formula for $\mathcal{G}$:

\begin{equation}
\mathcal{G}=\mathcal{G}_{0}+\mathcal{G}_{0}\cdot\delta\mathcal{D}\cdot\mathcal{G}.
\end{equation}

We substitute the matrix representation of $\delta\mathcal{D}$ given
above and iterate to obtain an explicit formula for $\mathcal{G}$
in terms of an infinite series:

\begin{equation}
\mathcal{G}=\mathcal{G}_{0}+\left(1+A+A^{2}+\ldots\right)G_{0}|x\rangle\hat{\boldsymbol{r}}_{i_{0}j_{0}}\otimes\hat{\boldsymbol{r}}_{i_{0}j_{0}}\langle y|\mathcal{G}_{0},
\end{equation}

\noindent where we have defined the constant $A=\hat{\boldsymbol{r}}_{i_{0}j_{0}}\langle y|\mathcal{D}_{0}^{-1}|x\rangle\hat{\boldsymbol{r}}_{i_{0}j_{0}}$.
We then sum the geometric series to obtain:

\begin{equation}
\mathcal{G}=\mathcal{G}_{0}+\frac{\mathcal{G}_{0}|x\rangle\cdot\hat{\boldsymbol{r}}_{i_{0}j_{0}}\otimes\hat{\boldsymbol{r}}_{i_{0}j_{0}}\cdot\langle y|\mathcal{G}_{0}}{1-\hat{\boldsymbol{r}}_{i_{0}j_{0}}\langle y|\mathcal{G}_{0}|x\rangle\hat{\boldsymbol{r}}_{i_{0}j_{0}}}.
\end{equation}

\noindent This is an explicit formula for the lattice Green's tensor
of the perturbed network in terms of the lattice Green's tensor
of the unperturbed network. The local stiffness measured between vertices
$i$ and $j$ is now

\begin{equation}
\tilde{k}_{\mathrm{loc}}=\frac{f_{0}}{u_{0}+\delta u},
\end{equation}

\noindent with the perturbation $\delta u$ to the measured deformation
given by:
\begin{equation}
\frac{\delta u}{f_{0}}=\frac{\left[\hat{\boldsymbol{r}}_{ii_{0}}\cdot\mathcal{G}_{0}(i,i_{0})\cdot\hat{\boldsymbol{r}}_{i_{0}j_{0}}-\hat{\boldsymbol{r}}_{ii_{0}}\cdot\mathcal{G}_{0}(i,j_{0})\cdot\hat{\boldsymbol{r}}_{i_{0}j_{0}}-\hat{\boldsymbol{r}}_{ji_{0}}\cdot\mathcal{G}_{0}(j,i_{0})\cdot\hat{\boldsymbol{r}}_{i_{0}j_{0}}+\hat{\boldsymbol{r}}_{ji_{0}}\cdot\mathcal{G}_{0}(j,j_{0})\cdot\hat{\boldsymbol{r}}_{i_{0}j_{0}}\right]^{2}}{1-2\hat{\boldsymbol{r}}_{i_{0}j_{0}}\cdot\left[\mathcal{G}_{0}(i_{0},i_{0})-\mathcal{G}_{0}(i_{0},j_{0})\right]\cdot\hat{\boldsymbol{r}}_{i_{0}j_{0}}}.
\end{equation}

\noindent Using this exact formula for the local stiffness in the
vicinity of a single removed fiber, we now approximate the stiffness loss, defined as the relative change in local stiffness upon removing
the bond $i_{0}j_{0}$ at a distance $R$ (see Fig. S3):

\begin{equation}
\Delta=\left({k}{}_{\textrm{loc}}-\tilde{k}_{\textrm{loc}}\right)/k_{\textrm{loc}}.
\end{equation}

When the location of the missing fiber is far away from the probe,
i.e. $R\gg d$, the perturbation is small and scales as:

\begin{equation}
\Delta\sim\delta u.
\end{equation}

\noindent At these long distances, the lattice Green's tensor approaches
the continuum elastic Green's tensor, which is given by
the following (41):

\begin{equation}
\mathcal{G}_{ik}(\boldsymbol{r})=\frac{1}{8\pi E}\left(\frac{1+\nu}{1-\nu}\right)\left[(3-4\nu)\delta_{ik}+n_{i}n_{k}\right]\frac{1}{r},
\end{equation}

\noindent where $E$ is the Young's modulus, $\nu$ is the Poisson's
ratio, $\delta_{ik}$ is the Kronecker delta function, $n_{i}$ is
a unit vector parallel to the position vector $\boldsymbol{r}$, and
$r$ is the magnitude of the position vector. Upon substituting the continuum
elastic Green's tensor in place of the lattice Green's tensor and
expanding the perturbation to third order in $1/R$, we find the stiffness loss scales as

\begin{equation}
\boxed{\Delta\sim\frac{1}{R^{2D}}}
\end{equation}

\noindent in $D$ dimensions. Comparing this universal result to the average stiffness loss measured
for disordered networks (Fig. 1\emph{C},\emph{D}, Fig. S1\emph{F}, and Fig. S2\emph{B}, \emph{Insets}), we find that the observed stiffness loss
is consistent with the $1/R^{2D}$ scaling prediction for all types of networks considered.

%FigS77
\begin{figure*}[t!]
\centerline{
\includegraphics[scale=0.25]{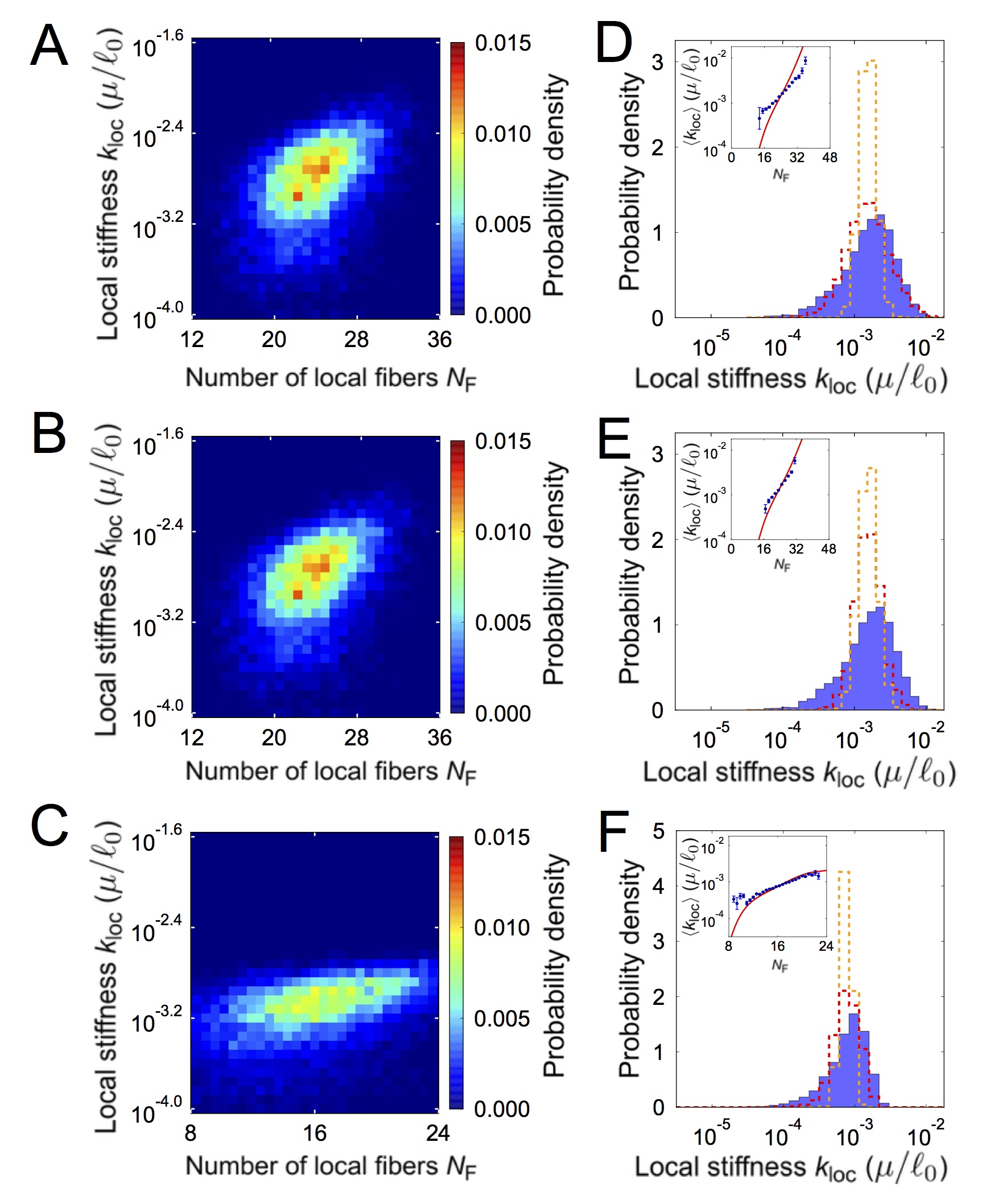}
}
\caption{\label{fig:F77}
Dependence of local stiffness on local structure for disordered lattice networks. (\emph{A}-\emph{C}) Joint distribution of local stiffness and number of local fibers $N_F$, defined as a sum of local bonds, for (\emph{A}) FCC network with bond weight $\Delta_{ij}=1$ below $\tilde{\xi}=\ell_0$ and decaying as $1/R_{ij}^{2D}$ beyond, (\emph{B}) FCC network with bond weight $\Delta_{ij}=1$ below $\tilde{\xi}=2\ell_0$ and zero beyond, and (\emph{C}) simple-cubic network with bond weight $\Delta_{ij}=1$ below $\tilde{\xi}=\ell_0$ and decaying as $1/R_{ij}^{2D}$ beyond. (\emph{D}-\emph{F}) Distribution of local stiffnesses for (\emph{D} and \emph{E}) FCC network ($d = \sqrt{5} \ell_0$, bond occupancy $p=0.3$, and bending modulus $\kappa=10^{-5} \mu \ell_0^2$ where $\mu$ is the stretching modulus), and (\emph{F}) simple-cubic network ($d = \sqrt{6} \ell_0$, bond occupancy $p=0.6$, and bending modulus $\kappa=10^{-5} \mu \ell_0^2$). Dashed lines show estimated distributions (assuming strong locality as in \emph{A}-\emph{C} red, weak locality orange, where weak locality is defined by bond weights that decay as $1/R^D$ beyond the cutoff $\tilde{\xi}$ (or zero beyond the hard cutoff $\tilde{\xi}=4\ell_0$ for \emph{B}). \emph{Insets}: geometric mean of marginal stiffness distribution $k_{\mathrm{loc}}(N_F)$ versus $N_F$. Error bars are given by the geometric standard deviations divided by the square root of the number of data points; red curves shows the macroscopic shear modulus versus the macroscopic average of the number of local fibers.
}
\end{figure*}

%FigS78
\begin{figure*}[t!]
\centerline{
\includegraphics[scale=0.25]{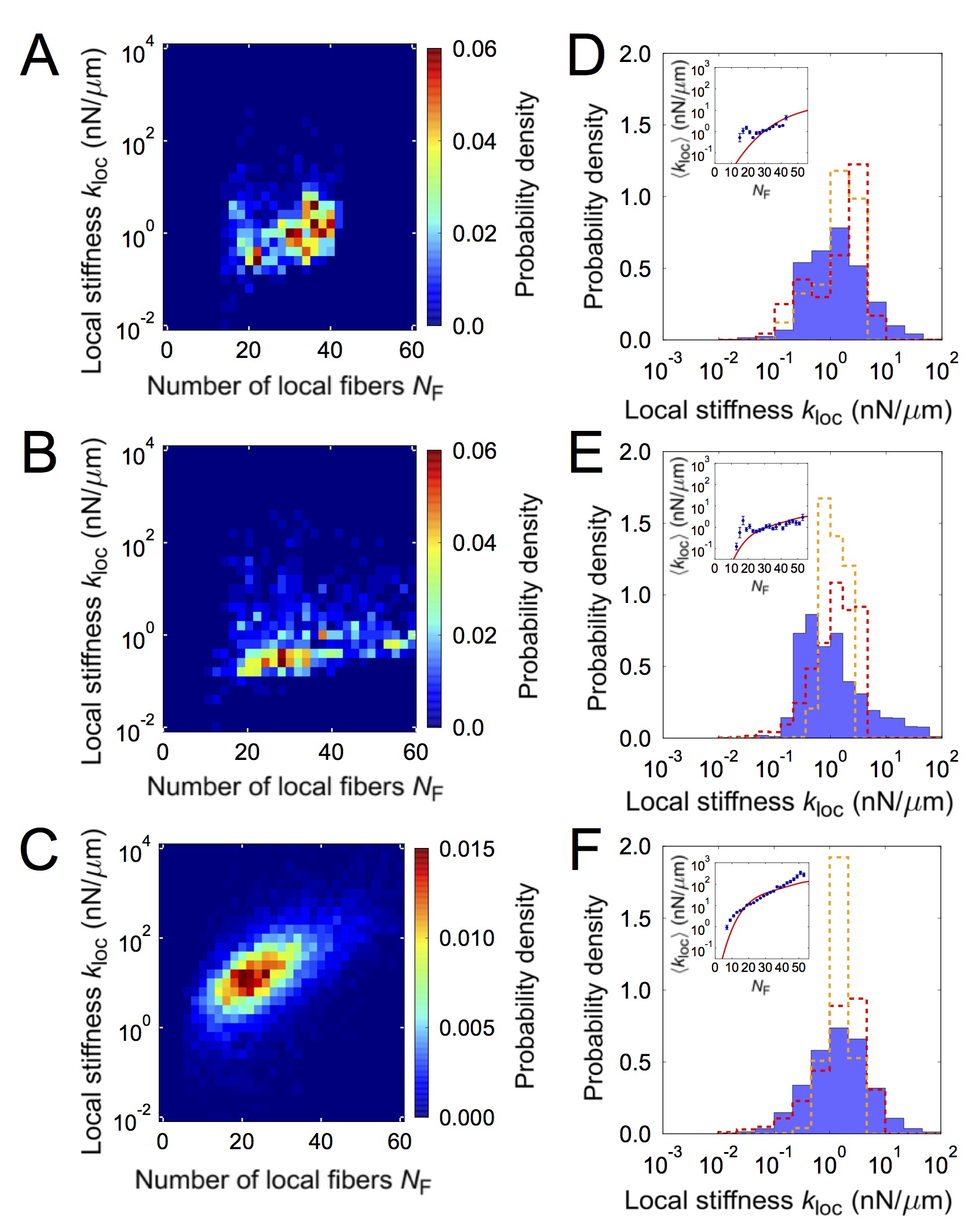}
}
\caption{\label{fig:F78}
Dependence of local stiffness on local structure for experimental and RGG networks. (\emph{A}-\emph{C}) Joint distribution of local stiffness and number of local fibers $N_F$, defined as a sum of local bonds, for (A) collagen network with bond weight $\Delta_{ij}=1$ below $\tilde{\xi}=2\xi$ ($\xi \simeq 6.5\ \mu$m for collagen) and decaying as $1/R_{ij}^{2D}$ beyond, (\emph{B}) fibrin network with bond weight $\Delta_{ij}=1$ below $\tilde{\xi}=1.5\xi$ ($\xi \simeq 6.7\ \mu$m for fibrin) and decaying as $1/R_{ij}^{2D}$ beyond, and (\emph{C}) RGG network with bond weight $\Delta_{ij}=1$ below $\tilde{\xi}=1.5\xi$ ($\xi \simeq 6.7\ \mu$m for RGG) and decaying as $1/R_{ij}^{2D}$ beyond. (\emph{D}-\emph{F}) Distribution of local stiffnesses for (\emph{D}) collagen network network ($d \sim 15 \mu$m), (\emph{E}) fibrin network ($d \sim 15 \mu$m), and (\emph{F}) RGG network ($d \sim 15 \mu$m). Dashed lines show estimated distributions (assuming strong locality as in \emph{A}-\emph{C} red, weak locality orange, where weak locality is defined by bond weights that decay as $1/R^D$ beyond the cutoff $\tilde{\xi}$). \emph{Insets}: geometric mean of marginal stiffness distribution $k_{\mathrm{loc}}(N_F)$ versus $N_F$. Error bars are given by the geometric standard deviations divided by the square root of the number of data points; red curves shows the macroscopic shear modulus versus the macroscopic average of the number of local fibers.
}
\end{figure*}

\subsection{Numerical estimate of the local stiffness distribution.}

The rapid decay of stiffness loss suggests that the local stiffness largely depends on a very small number of proximal fibers. To quantify this dependence, we estimate the local stiffness distribution by considering the number of local fibers $N_F$, given by:

\begin{equation}
N_{F}=\sum_{\langle ij\rangle} \Delta_{ij},
\end{equation}

\noindent where $\Delta$ is a rapidly decaying function that measures
the influence of a fiber based on the distance $R_{ij}$ from the center
of the probe to the center of the removed fiber:

\begin{equation}
\Delta_{ij}=\begin{cases}
1 & R_{ij}<\tilde{\xi},\\
\left(\frac{\tilde{\xi}}{R_{ij}}\right)^{x} & R>\tilde{\xi}.
\end{cases}
\end{equation}

\noindent Here, for simplicity we characterize locality by the strength $x$ of the power-law decay beyond a short-range cutoff $\tilde{\xi}$, which we have introduced to eliminate unphysical divergences. Our introduction of a variable exponent $x$ will allow us to later check how the variance of $N_F$ (and consequently our estimate for the local stiffness distribution) depends on the strength of locality. For strong locality, i.e. equal to the rapid $1/R^6$ decay of stiffness loss $\Delta$ in three dimensions, $k_{\mathrm{loc}}$ is strongly correlated with $N_F$ for all types of networks studied (Fig. S4\emph{A}-\emph{C} and Fig. S5\emph{A}-\emph{C}, \emph{Insets}). This correlation suggests that $N_F$ explains much of the variance of local stiffness. However, the geometric mean of the marginal distribution of local stiffnesses $\langle k_{\mathrm{loc}}(N_F) \rangle$ scales rapidly and nonlinearly with $N_F$. Consequently, the geometric standard deviation of $N_F$ is less than half of that of $k_{\mathrm{loc}}$. This underestimation occurs because fibers have inherently cooperative contributions to the local stiffness that are not captured by the weighted sum in Eq. (S38), which implicitly assumes that fibers contribute to $k_{\mathrm{loc}}$ independently.

To understand how the strength of the collective effects scales with fiber density, we considered the macroscopic shear modulus as a function of average coordination number $z$, or equivalently $\langle N_F \rangle$ for a fixed density of vertices. We found that the scaling of the shear modulus with $z$ was consistent with the scaling of $\langle k_{\mathrm{loc}}(N_F) \rangle$ with $N_F$ (Fig. S4\emph{D}-\emph{F}, Fig. S5\emph{D}-\emph{F}). We therefore estimated the local stiffness distribution for all types of networks by transforming the numerically-obtained distribution of $N_F$ by the functional dependence of the corresponding shear modulus $G(N_F)$ for a network with average number of fibers equal to $\langle N_F \rangle$, which resulted in a geometric standard deviation given by:

\begin{equation}
\log \sigma_{\mathrm{loc}} = \sqrt { \int \left[ \log G(N_F) - \langle \log G(N_F) \rangle \right]^2 P(N_F) dN_F },
\end{equation}

\noindent where

\begin{equation}
\langle \log G(N_F) \rangle = \int  \log G(N_F) P(N_F) dN_F.
\end{equation}

For strong locality, which for concreteness we consider throughout our analysis to be $x=6$ (equal to the rapid $1/R^{2D}$ scaling of the stiffness loss), the estimator captures the majority of the geometric standard deviation for all types of networks we studied (Table 1, Fig. 1\emph{C},\emph{D}, Fig. S1\emph{F}, Fig. S2\emph{B}, and Fig. S4\emph{F}). Furthermore, the estimate does not depend sensitively on the details of the form of the weighting function, as long as the decay is rapid enough (Fig. S4\emph{D},\emph{E}). Importantly, this estimator captures the large quantitative discrepancy between the width of the experimental and RGG networks versus that of the lattice networks (Table 1).

\begin{center}
  \begin{tabular}{  c  | c  c  c  c  c }
    & Collagen & Fibrin & RGG & FCC & SC \\ \hline
    $\sigma_{\mathrm{loc}}$ & 0.54 & 0.63 & 0.56 & 0.37 & 0.29 \\ 
    $\sigma_{\mathrm{th}}$ & 0.49 & 0.40 & 0.49 & 0.30 & 0.19 \\ 
  \end{tabular}
  \\[5 pt]
{\textbf{Table 1:} Numerically-simulated and theoretically-estimated geometric standard deviations for all types of networks studied.}
\end{center}

However, for weak locality, which we consider throughout our analysis as $x=3$, the estimated geometric standard deviation is significantly smaller than the observed value (Fig. 1\emph{C},\emph{D}, Fig. S1\emph{F}, Fig. S2\emph{B}, and Fig. S4\emph{F}).

\clearpage

\section{Continuum theory for the mean stiffness}

%FigS3
\begin{figure*}[t!]
\centerline{
\includegraphics[width=\columnwidth]{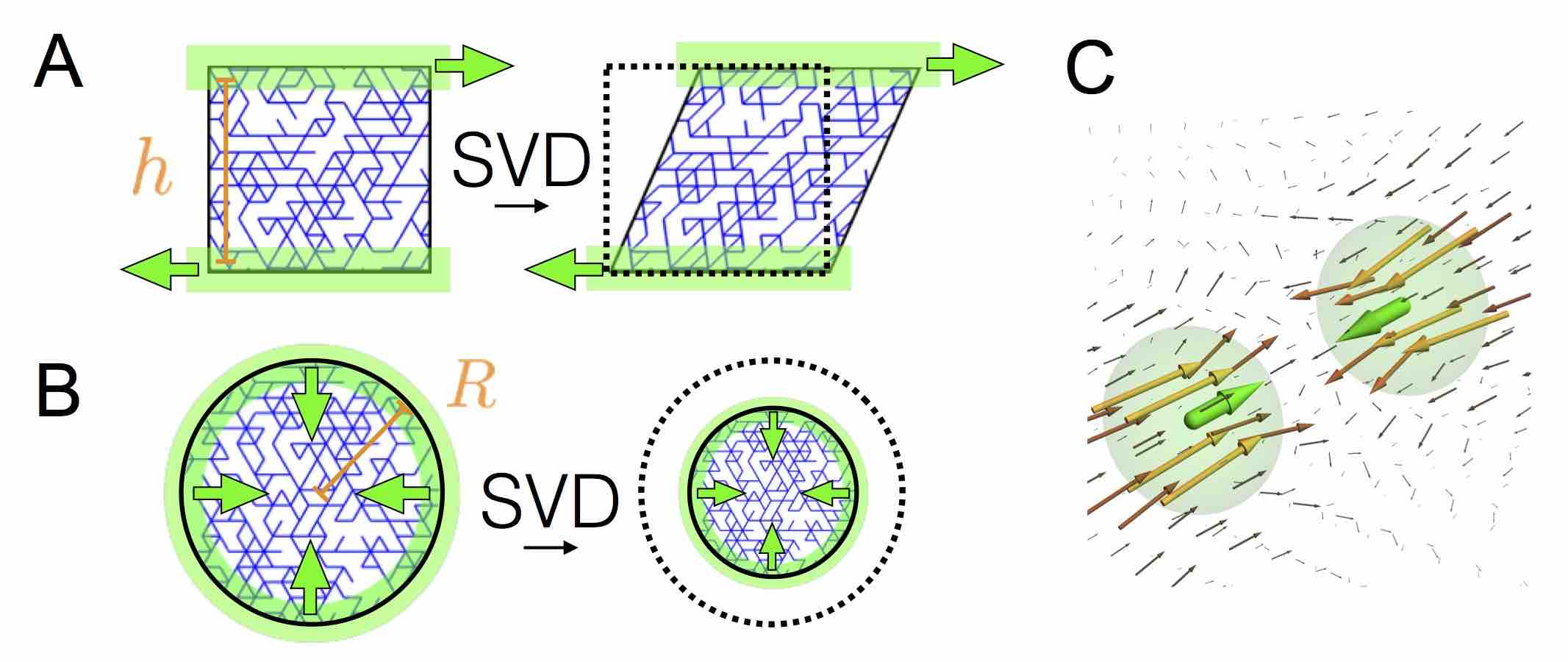}
}
\caption{\label{fig:F3}
(\emph{A} and \emph{B}) Schematic illustration of the procedure used to compute the elastic moduli used in the continuum theory. (\emph{A}) For the shear modulus, SVD is used to calculate the deformation response to shear forces applied to nodes within buffer zones at the top and bottom of a cubical region of network. The shear modulus is defined as $G=\frac{F/A}{\bar{\gamma}}$, where $F$ is the sum of the magnitudes of all the forces applied to the network, $A$ is the area of the buffer region, and $\bar{\gamma}$ is the average shear strain obtained from the least-squares linear fit of the deformation profile in the direction of the applied force. (\emph{B}) For the bulk modulus, SVD is used to simulate the deformation response to compressive forces applied to nodes within a buffer zone at the boundary of a spherical region of network. The bulk modulus is defined as $B=F/(24\pi R^2 \bar{\epsilon})$ where $F$ is the sum of the magnitudes of all the forces applied to the network, $R$ is the distance from the center of the network to the center of the buffer zone, and $\bar{\epsilon}$ is the average radial strain of the nodes within the bulk of the network. (\emph{C}) Schematic illustration of the continuum model for mean stiffness of a network. A local stiffness probe is modeled by two force-monopoles, separated by a distance $d$, that apply a contractile force to an isotropic continuum elastic material with moduli calculated from the network. The local stiffness $k_{\mathrm{loc}}$ in the continuum model is defined as the sum of the average deformation within spheres of radius $\xi$ centered at the points of force application, where $\xi$ is a short-range cutoff equal to the mesh size, defined for the experimental and RGG networks as the radius of a sphere containing the average volume per node, and for the disordered lattice networks as the lattice spacing.
}
\end{figure*}

In this section, we present a simple model for the mean stiffness
measured by a local stiffness probe interacting with an isotropic
continuum elastic material in three dimensions.

\subsection{Effective continuum elastic-material model.}

To gain insight into the average mechanical response of a disordered
fiber network to a local force-dipole, we approximate the network
as an isotropic continuum elastic material (41). The response of such an isotropic continuum
elastic material can be completely characterized by two macroscopic
moduli that appear in the elastic free energy:

\begin{equation}
F=\frac{1}{2}\lambda \epsilon_{ii}\epsilon_{jj} + G \epsilon_{ij}\epsilon_{ij},
\end{equation}

\noindent where $\epsilon_{ij}= 1/2 \left( {\partial u_i}/{\partial x_j} + {\partial u_j}/{\partial x_i} \right)$ is the strain tensor for the deformation field $u_i$, $\lambda$ is the first Lame coefficient, and $G$ is the shear modulus. To determine these elastic parameters of our model, we calculate the shear modulus $G$ and the bulk modulus $K=\lambda+({2}/{3})G$ of the network.

To compute the shear modulus of a network, we simulate the deformation
of a cubical region of network in response to a shear stress. We apply
a shear stress via force monopoles applied to all nodes within buffer
zones at the top and bottom of the network (see Fig. S6\emph{A}). The deformations
of the vertices are obtained by solving the force-balance equation
for static equilibrium using SVD (see \emph{Numerical procedure}). The shear modulus
is defined as

\begin{equation}
G=\frac{F/A}{\bar{\gamma}},
\end{equation}

\noindent where $F$ is the sum of the magnitudes of all the forces
applied to the network, $A$ is the area of the buffer region, and $\bar{\gamma}$
is the average shear strain obtained from the least-squares linear fit of the
deformation profile in the direction of the applied force.

To compute the bulk modulus, we simulate the deformation of a spherical
region of network in response to a uniform compression. We apply a
uniform compression via force-monopoles applied to nodes at the edge
of the network (see Fig. S6\emph{B}). The deformations of the vertices are
obtained as before by solving the force-balance equation for static
equilibrium using SVD. The bulk modulus is defined as

\begin{equation}
B=F/(24\pi R^{2}\bar{\epsilon}),
\end{equation}

\noindent where $F$ is the sum of the magnitudes of all the forces
applied to the network, $R$ is the distance from the center of the
network to the center of the buffer zone, and $\bar{\epsilon}$ is
the average radial strain of the nodes within the bulk of the network. Here, the geometric prefactor $\frac{1}{24 \pi}$ arises from the spherical geometry.

\subsection{Continuum stiffness-measuring force-dipole model.}

The local stiffness probe consists of two force-monopoles separated
by a distance $\boldsymbol{d}$ which exert a contractile force of
magnitude $f_{0}$ (see Figure S6\emph{C}):

\begin{equation}
\boldsymbol{f}(\boldsymbol{r})=f_{0}\left[\delta(\boldsymbol{r})-\delta(\boldsymbol{r}-\boldsymbol{d})\right]\hat{\boldsymbol{d}}.
\end{equation}

\noindent The deformation produced by each force-monopole is given
by the continuum elastic Green's tensor (Eq. S36). The continuum elastic Green's
tensor relates applied forces to the resulting deformations of the
medium as follows:

\begin{equation}
\boldsymbol{u}(\boldsymbol{r})=\int\mathcal{G}(\boldsymbol{r})\cdot\boldsymbol{f}(\boldsymbol{r})\ dV.
\end{equation}

\noindent Thus, the combination of both force-monopoles acting on
the medium results in a deformation $\boldsymbol{u}$ at each point
$\boldsymbol{r}$ within the medium:

\begin{equation}
\boldsymbol{u}(\boldsymbol{r})=f_{0}\left[\mathcal{G}(\boldsymbol{r})-\mathcal{G}(\boldsymbol{r}-\boldsymbol{d})\right]\cdot\hat{\boldsymbol{d}}.
\end{equation}

We would like to define a local stiffness probe that measures the
sum of the deformation $u$ at each monopole along the direction of
the probe:

\begin{equation}
u=\boldsymbol{u}(0)\cdot\hat{\boldsymbol{d}}-\boldsymbol{u}(\boldsymbol{d})\cdot\hat{\boldsymbol{d}}=2\boldsymbol{u}(0)\cdot\hat{\boldsymbol{d}},
\end{equation}

\noindent however, in contrast to the network model, here in the continuum model, the force-monopoles
induce diverging deformations at the points where the forces are applied.

To account for the unphysical divergences, we define the measured
deformation at each point of force application as the average deformation
within a spherical cutoff region of radius $\xi$ centered around
the point, where $\xi$ is set equal to the mesh size. Intuitively, this choice of cutoff results in a modeled deformation that corresponds to a coarse-grained description of the deformation of a single vertex in the network. The total measured deformation is then given by:

\begin{equation}
\bar{u}=\frac{3}{4\pi\xi^{3}}\int_{r\le\xi}u\ d\boldsymbol{r}.
\end{equation}

\noindent We evaluate the integral assuming $d>\xi$ and find
the local stiffness measured by the ``continuum dipole'' is:

\begin{equation}
k_{\mathrm{loc}}=\frac{f_{0}}{\bar{u}}=\left(\frac{3a_{0}}{\xi}-\frac{a_{0}}{d}+\frac{b_{0}\xi^{2}}{d^{3}}\right)^{-1},
\end{equation}

\noindent where we have defined two constants that depend on the elastic
moduli of the network:

\begin{equation}
a_{0}=\frac{1+\nu}{\pi E},
\end{equation}

\begin{equation}
b_{0}=\frac{1}{10\pi E}\left(\frac{1+\nu}{1-\nu}\right).
\end{equation}

The continuum theory results in monotonically decreasing stiffness
as a function of probe length, with stiffness asymptotically approaching a constant
value for $d\gg\xi$. This occurs because the measured deformations
are produced by two force-monopoles, which add in a linear fashion. Since
the self-deformation $u_{0}=3a_{0}/2\xi$ of each monopole remains
constant, and the contribution of the far monopole at $d$ decays
to zero for $d\gg\xi$, the local stiffness saturates for large
probe lengths, i.e. it approaches a constant value determined by
the combined deformations of two non-interacting monopoles.

To study the dynamic range of local stiffness as a function of probe length, we consider the fold-stiffening, defined as the
ratio $\langle k_{\mathrm{loc}} \rangle /\langle k_{\mathrm{\infty}}\rangle$ of the geometric mean stiffness to the geometric mean stiffness for asymptotically long
probes (i.e. $d\rightarrow\infty$):

\begin{equation}
\frac{ \langle k_{\mathrm{loc}} \rangle }{\langle k_{\mathrm{\infty}}\rangle}=\left(1-\frac{\xi}{3d}+\frac{b_{0}\xi^{3}}{3a_{0}d^{3}}\right)^{-1}.
\end{equation}

This fold-stiffening predicted from continuum theory is roughly consistent
with the fold-stiffening for all types of discrete networks studied when the probe length is at least five times the mesh size (Fig. S7). Despite this match in the fold-stiffening at long probe lengths, the observed local stiffness for short probe lengths is always much stiffer than the continuum prediction for the experimental and RGG networks (Fig. S7\emph{A}-\emph{C}).
This discrepancy occurs because the continuum theory does not adequately describe the deformation produced by a monopole at short distances for these networks. This results in a fixed difference between the force-monopole self-deformation predicted by continuum theory and the average deformation produced by a force-monopole applied to the discrete networks.

\subsection{The effect of long fibers.}

%FigS4
\begin{figure*}[t!]
\centerline{
\includegraphics[scale=0.2]{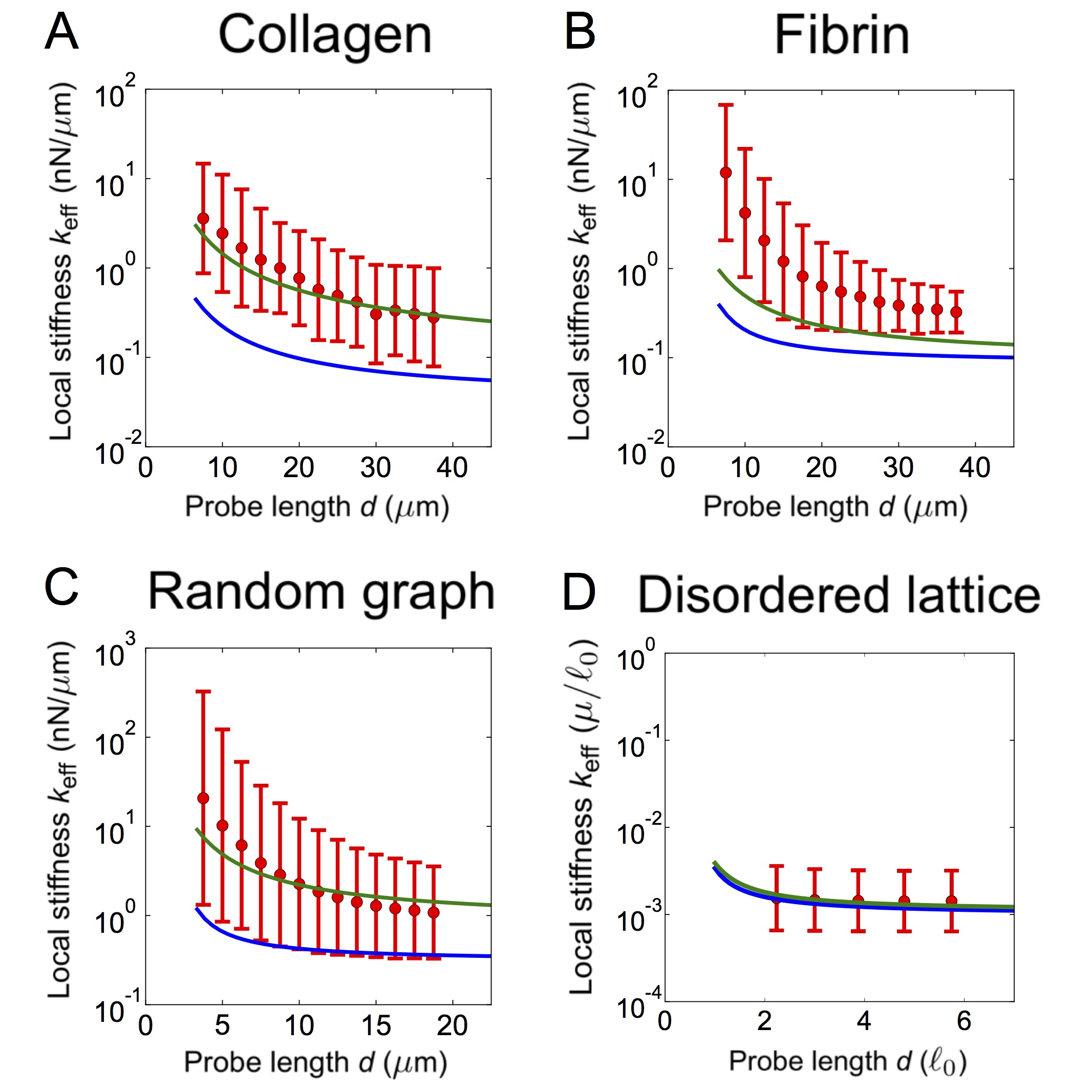}
}
\caption{\label{fig:F4}
(\emph{A}-\emph{D}) Local stiffness $k_{\mathrm{loc}}$ versus probe length $d$ for (\emph{A}) experimental collagen network, (\emph{B}) experimental fibrin network, (\emph{C}) RGG network, and (\emph{D}) FCC network, showing geometric mean $\langle k_{\mathrm{loc}} \rangle$ and geometric standard deviation $\sigma_{\mathrm{loc}}$ of local stiffness (red); solid curves from conventional continuum elastic theory (blue) and gradient elastic theory (green).
}
\end{figure*}

Below around five times the mesh size, the fold-stiffening measured for the experimental and RGG networks begins to deviate rapidly from the continuum prediction (Fig. S7\emph{A}-\emph{C}). In contrast, the continuum theory seems to capture the fold-stiffening of the FCC network, which is fairly constant, even down to probe lengths close to the lattice spacing (Fig. S7\emph{D}). What determines the length scale of the enhanced fold-stiffening we observed for the experimental and RGG networks? In the previous sections, we saw that for fixed probe length, the range of stiffnesses is determined in part by the strong, nonlinear dependence of the marginal distribution of local stiffnesses on network connectivity, with the strongest dependence occurring close to the stretching and bending transitions. Does the scaling of the marginal distribution of local stiffnesses also depend on probe length? If so, the large fold-stiffening we observed for the experimental and RGG networks could occur due to the proximity to an elastic transition.

To determine how network connectivity impacts the fold-stiffening, we varied the average coordination number $z$ for a fixed ratio of the bending modulus $\kappa$ to the stretching modulus $\mu$. We found that for the RGG network, the dependence of fold-stiffening on probe length becomes much stronger as the average coordination number is tuned to bring the networks closer to an elastic transition (Fig. S8). However, even far away from the transitions, there is still a large fold-stiffening at short probe lengths. Since only the experimental and RGG networks have this large fold-stiffening away from the transitions, this property
must arise from intrinsic features of the networks that are not present for the FCC network. One feature of the network architecture that explicitly introduces a length scale is the distribution of fiber lengths (Fig. S9).

%FigS99
\begin{figure*}[t!]
\centerline{
\includegraphics[scale=0.12]{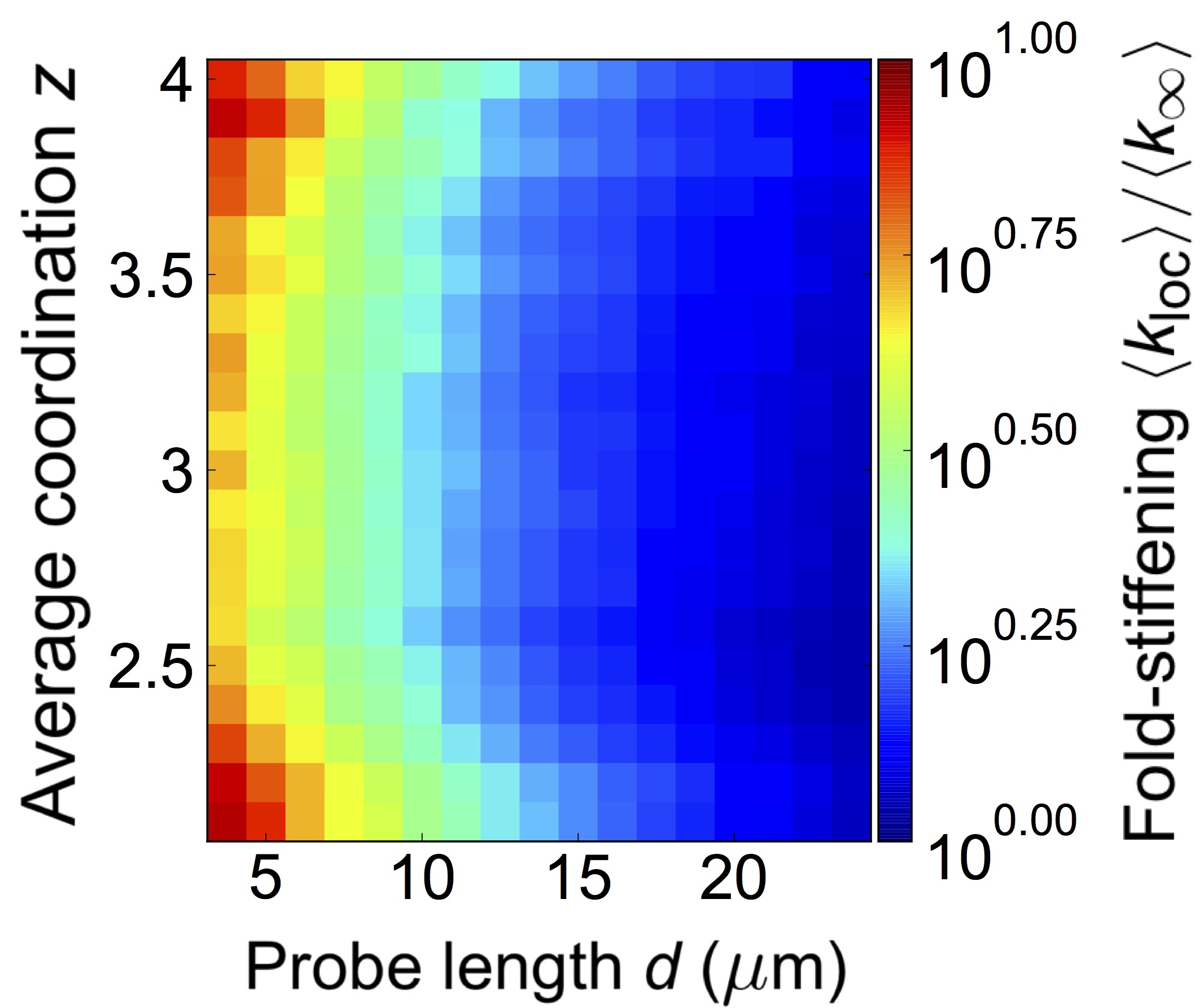}
}
\caption{\label{fig:F99}
Fold-stiffening $\langle k_{\mathrm{loc}} \rangle / \langle k_{\mathrm{\infty}} \rangle$ versus probe length $d$ for RGG network over a range of average coordination number $z$.}
\end{figure*}

To predict the probe length for which we expect the crossover to short-range stiffening to occur due to the presence of polydisperse fibers, we consider the distribution of fiber lengths for each network. For the disordered lattice networks, we define fiber length as the number of bonds in a filament that are consecutive coaxial segments uninterrupted by a non-coaxial connected segment. We approximate the distribution of fiber lengths $P(\ell)$ as proportional to the probability of starting from a fiber endpoint and observing $\ell$ consecutive bonds followed by either a cross-linking (i.e. non-coaxial) bond or the absence of a bond, yielding the following distribution:

\begin{equation}
P(\ell)\propto\left[p(1-p)^{z-2}\right]^{\ell},
\end{equation}

\noindent where $z$ is the coordination number (6 for the SC lattice and 12 for an FCC lattice). Thus, we find an exponentially decaying distribution with a decay length $L$ given by:

\begin{equation}
L=\frac{-1}{\log\left(p(1-p)^{z-2}\right)}.
\end{equation}
 
\noindent We see that for the lattice networks, the distribution of fibers decays exponentially. For the FCC network at connectivity $p=0.3$, the decay length is around $L=0.2$. This is approximately an order of magnitude shorter than the minimum probe length we considered ($d\gtrsim2$), which explains why we did not observe the fold-stiffening to deviate from the continuum prediction away from the elastic transitions.

%FigS9999
\begin{figure*}[t!]
\centerline{
\includegraphics[scale=0.2]{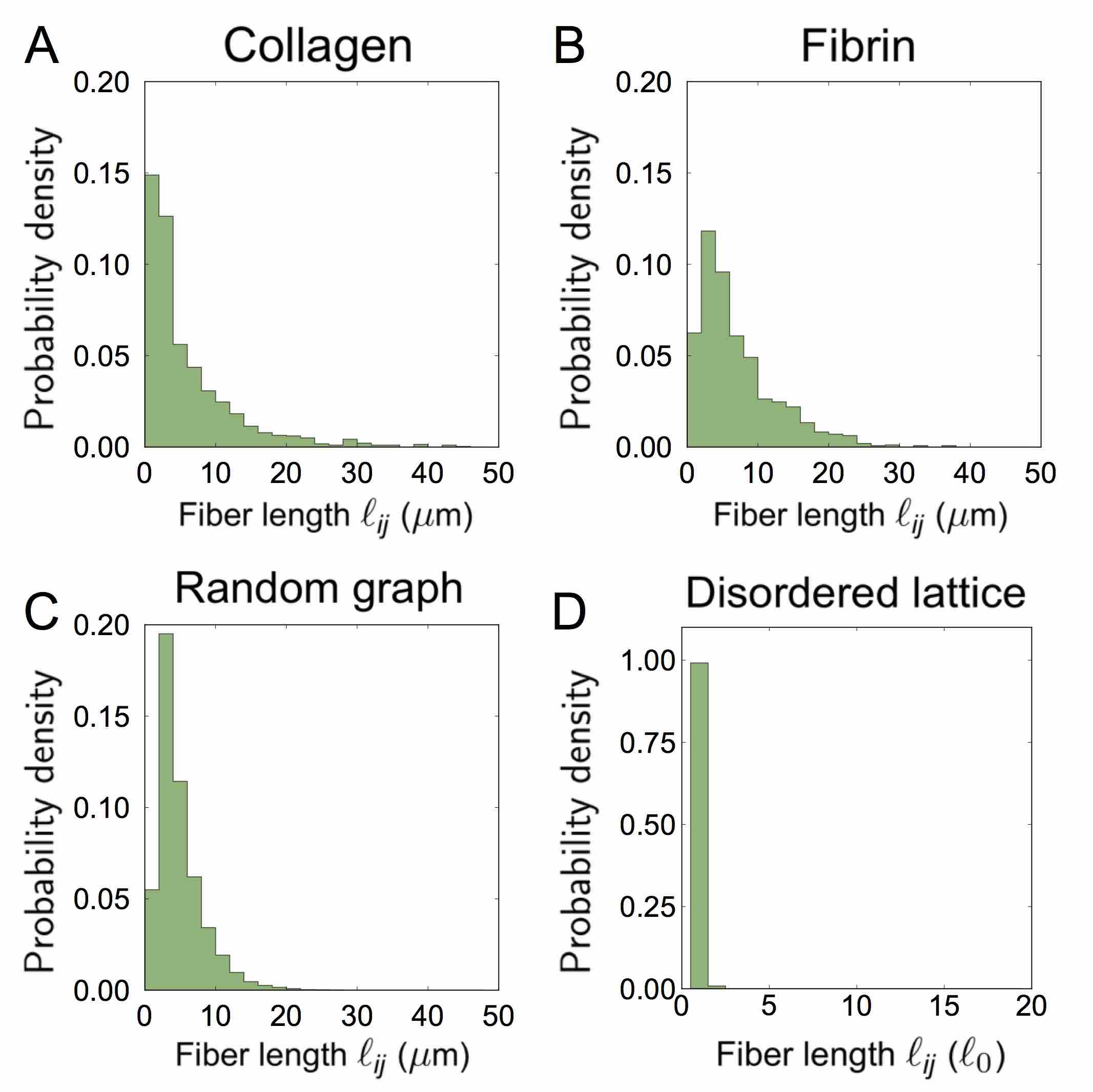}
}
\caption{\label{fig:F99}
(\emph{A}-{D}) Distribution of fiber lengths for (\emph{A}) experimental fibrin network, (\emph{B}) experimental collagen network, (\emph{C}) RGG network, and (\emph{D}) FCC network.}
\end{figure*}

In contrast to the relatively monodisperse fiber-length distribution for the FCC network, the experimental and RGG networks contain long fibers that connect vertices separated by distances that are well beyond the mesh size. Beyond the mesh size $\xi$, the fiber-length distribution decays roughly exponentially with a length scale approximately equal to the mesh size $L\simeq\xi$. Thus, we expect the long fibers to modify the fold-stiffening out to intermediate probe lengths. Could this effect account for the larger fold-stiffening we observed for the experimental and RGG networks?

The long-ranged structure in the network, e.g. fibers that are several times longer than the probe lengths we considered, precludes us from considering the free energy in the hydrodynamic limit. Thus, to properly describe these networks at these intermediate length scales, we should also consider terms in the free energy that consist, for instance, of gradients of the strain tensor. Although many higher-order terms could contribute at such scales, for simplicity we consider an extension that contains a single additional term (42). Specifically, for each term in the original elastic energy which contains the squared-sum of the diagonal components of the strain tensor $\epsilon_{ii}$ or the squared-sum of the individual components $\epsilon_{ik}$, we also add the vector product of the gradient of each term with itself:

\begin{equation}
\tilde{F} =\frac{1}{2}\lambda \epsilon_{ii}\epsilon_{jj} + G \epsilon_{ij}\epsilon_{ij} + L^2 \left( \frac{1}{2}\lambda \epsilon_{ii,k}\epsilon_{jj,k} + G \epsilon_{ij,k}\epsilon_{ij,k} \right),
\end{equation}

\noindent where $\epsilon_{ij,k}$ is the partial derivative of $\epsilon_{ij}$ with respect to the coordinate represented by index $k$, and $L$ is an additional material parameter that corresponds to the extent of the microscopic non-locality. By setting $L$ equal to the decay length obtained from the fiber length distributions of the networks, we show that this gradient elasticity theory captures the essential features shown in our simulations. The Green's tensor for this non-local continuum theory is given by:

\begin{equation}
\mathcal{G}_{ik}(\boldsymbol{r})=\frac{1}{32 \pi G (1-\nu)} \left[\Psi(r)\delta_{ik}+X(r)n_{i}n_{k}\right],
\end{equation}

\noindent where, as in (42), we have defined the two functions

\begin{equation}
\Psi(r)= \frac{2}{r} \left \{ (3-4\nu)\left(1-e^{-r/L} \right) + \frac{1}{r^2} \left[ 2L^2 - (r^2 + 2Lr + 2L^2)e^{-r/L} \right] \right \},
\end{equation}

\begin{equation}
X(r) = \frac{2}{r} \left[ \left( 1-\frac{6L^2}{r^2} \right) + \left( 2 + \frac{6L}{r} + \frac{6L^2}{r^2} \right) e^{-r/L} \right].
\end{equation}

\noindent Incorporating the additional effect of the strain gradient into the continuum prediction for the experimental and RGG networks results in a fold-stiffening that is approximately a decade at the mesh size $\xi$, which is significantly larger than the amount predicted by conventional continuum elasticity and captures a majority of the anomalously large fold-stiffening observed for the experimental and RGG networks (Fig. S7). In contrast, for the FCC network, the gradient prediction is nearly identical to the conventional continuum prediction, due to the much smaller value of $L$ relative to $\xi$. These results suggest that accounting for the inherent non-locality from highly polydisperse fibers is crucial to describe the response of biopolymer networks at the cellular scale.

\clearpage

\section{Generalized stiffness inference}

How accurately can cells infer the \emph{global} stiffness properties of their surroundings based solely on \emph{local} measurements? As we showed above, the global properties are closely related to the overall behavior of the local stiffness distribution. Specifically, the median stiffness tracks the scaling with network connectivity of the macroscopic response. This suggests that a suitable average of several samples of local stiffness could provide useful information about the bulk properties of the network.

To quantify the accuracy of inference, we considered an idealized sampling process that consists of averaging $N$ local stiffness measurements obtained within prolate spheroids. For a given spheroid, the inferred stiffness is given by the geometric mean of $N$ measurements randomly chosen among all the probes whose centers are contained within its volume. We define the relative uncertainty of this estimate as the geometric standard deviation $\sigma_N$ of the resulting distribution of inferred stiffnesses. This uncertainty is minimized for independent samples, for which $\sigma_N$ is related to the logarithm of the geometric standard deviation $\log \sigma_{\mathrm{loc}} = \sqrt { \langle (\log k_{\mathrm{loc}} - \langle \log k_{\mathrm{loc}} \rangle)^2 \rangle }$ of the distribution of local stiffnesses as follows:

\begin{equation}
\log \sigma_{\mathrm{N}} =  \frac{ \log \sigma_{\mathrm{loc}} } {\sqrt{N}}.
\end{equation}

\noindent In general, however, stiffness measurements are correlated in space because nearby stiffness measurements depend on shared local structure. To quantify these correlations, we consider the two-point geometric correlation function $\textit{C}_{ij}$ for local stiffness measurements $k^{(i)}_{\mathrm{loc}}$ and $k^{(j)}_{\mathrm{loc}}$, defined as follows:

\begin{equation}
\log \left( \textit{C}_{ij} \right) \equiv \frac{\langle (\log k^{(i)}_{\mathrm{loc}} - \langle \log k_{\mathrm{loc}} \rangle)(\log k^{(j)}_{\mathrm{loc}} - \langle \log k_{\mathrm{loc}} \rangle) \rangle}{ (\log \sigma_{\mathrm{loc}})^2 }.
\end{equation}

We calculated the two-point geometric correlation function for the experimental and RGG networks and found $\textit{C}_{ij}$ that were consistent with a decaying exponential form (Fig. 4\emph{B}, lower \emph{Inset}, Fig. S10A, \emph{Inset}):

\begin{equation}
\textit{C}_{ij} \simeq \frac{C_0 e^{-|\vec{R}_{ij}| / \Lambda}}{ (\log \sigma_{\mathrm{loc}})^2  },
\end{equation}

\noindent where $\vec{R}_{ij}=\vec{R}_j-\vec{R}_i$ is the separation between the centers $\vec{R}_i$ of the stiffness probes, $\Lambda$ is the two-point stiffness correlation length, and $C_0$ is the zero-distance geometric covariance of local stiffnesses whose centers lie at the same point ($C_0$ is less than $1$ because the centers of distinct probes may coincide). The correlation length $\Lambda$ and the zero-distance geometric covariance $C_0$ are obtained by fitting to the modeled local stiffnesses. Due to the presence of these spatial correlations, the uncertainty of inference $\sigma_N$ will be larger than the uncertainty $\sigma_{\mathrm{loc}} / \sqrt{N}$ predicted by assuming samples of local stiffness are independent. Provided that we can approximate the uncertainty using only the two-point geometric correlation function, the geometric standard deviation $\sigma \left( \left\{ \vec{R}_1,   \vec{R}_2, \ldots \vec{R}_N \right\} \right)$ (where $\left\{ \vec{R}_i \right\}$ is the position of sample $i$ for each of the $N$ samples) is given by:

\begin{equation}
\log \sigma \left( \left\{ \vec{R}_i \right\} \right) =  \left( 1 + \sum_{i<j} \log \textit{C}_{ij} \right)^{1/2}  \frac { \log \sigma_{\mathrm{loc}} } {\sqrt{N}}.
\end{equation}

\noindent For our idealized sampling process, the distances between the samples vary because the probes are positioned randomly throughout the network. In this case, each configuration provides an additive contribution to the variance with a weight proportional to its probability. Since the probes are distributed homogeneously and isotropically throughout space, the overall variance of the inference is simply given by integrating the positions of the $N$ samples over the sampling volume and normalizing by $V^N$. Upon substituting the form of $\textit{C}_{ij}$ from above, we find an uncertainty $\sigma_N$ given by:

\begin{equation}
 \log \sigma_{\mathrm{N}} = \left(  \frac{   (\log \sigma_{\mathrm{loc}})^2  }{N} + \frac{N (N-1) C_0}{4V^2} \int_{\vec{R}_1 \in V} \int_{\vec{R}_2 \in V} e^{-|\vec{R}_1 - \vec{R}_2| / \Lambda } \ d\vec{R}_1 d\vec{R}_2 \right)^{1/2}.
 \end{equation}

\noindent For the simple case in which the sampling volume is a sphere of radius $R_0$, this becomes:

\begin{equation}
 \log \sigma_{\mathrm{N}} = \left(  \frac{   (\log \sigma_{\mathrm{loc}})^2  }{N} + N (N-1) C_0   \frac{3 \Lambda^3 \left(  
 4R_0^3 - 9 R_0^2 \Lambda + 15 \Lambda^3 - 3 e^{-2R_0/ \Lambda} (R_0+\Lambda)(2R_0^2 + 5 R_0 \Lambda + 5 \Lambda^2)
  \right) }{8R_0^6}   \right)^{1/2}.
 \end{equation}
  
\noindent We measured the uncertainty $\sigma_2$ of the two-sample stiffness inference process for the RGG network and confirmed that it matches the above direct calculation based on the two-point geometric correlation of local stiffnesses of the network (Fig. S10). This consistency demonstrates that the overall uncertainty of inference for sampling two stiffnesses within a volume can be characterized in terms of the underlying two-point correlations among stiffness measurements.

%FigS10
\begin{figure*}[t!]
\centerline{
\includegraphics[width=\columnwidth]{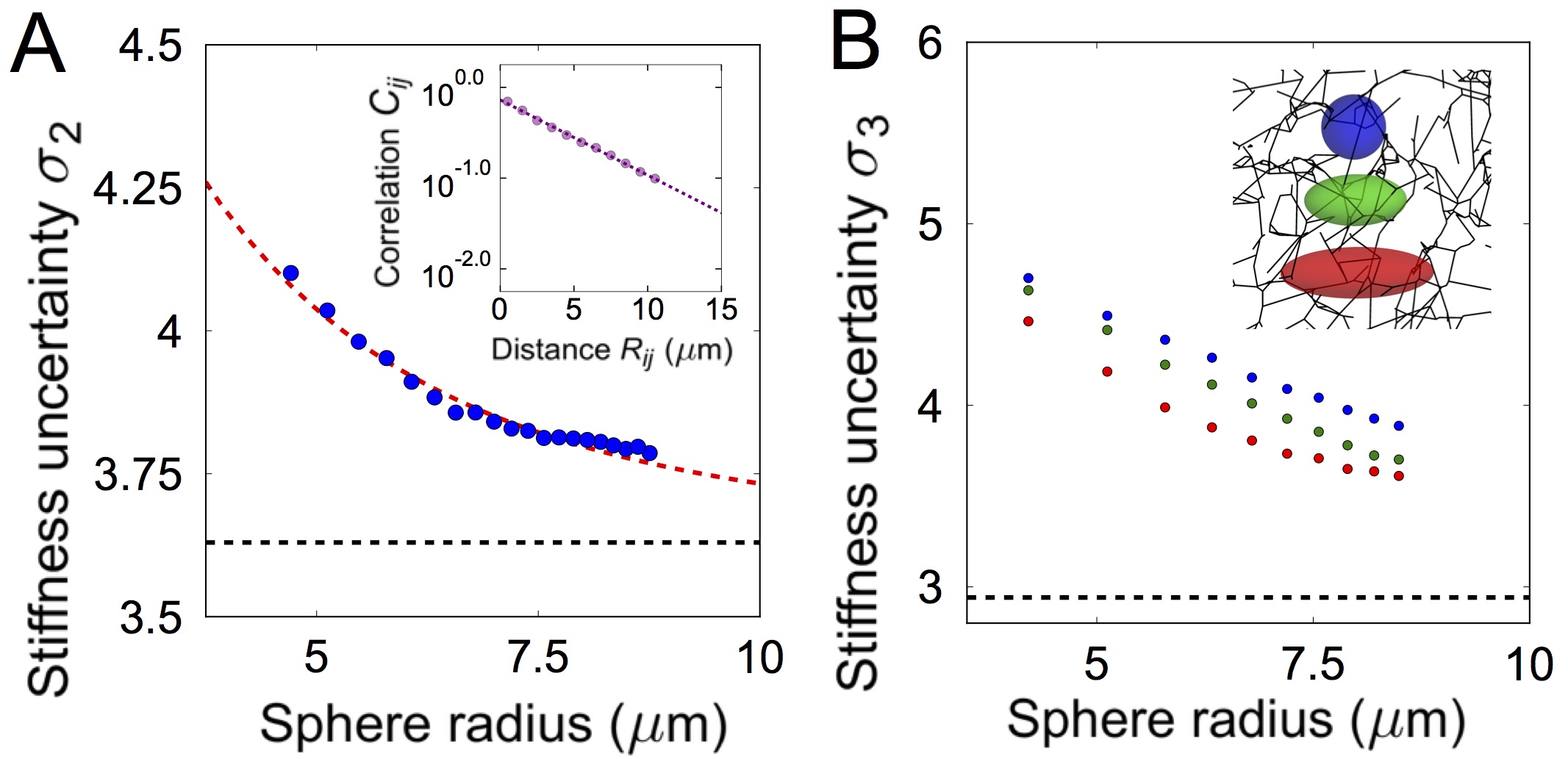}
}
\caption{\label{fig:F10}
Minimum uncertainty of stiffness inference for RGG network. (\emph{A}) Fractional uncertainty of stiffness inference from two measurements, defined as the geometric standard deviation $\sigma_{\mathrm{2}}$ of the geometric mean of a random sample of $N=2$ local stiffnesses measured by force dipoles whose centers lie within spheres versus radius of spheres. Dashed black line shows the geometric standard deviation $\sigma_{\mathrm{loc}}^{1/\sqrt{2}}$ of the geometric mean of $N=2$ independent measurements, dashed red line shows the uncertainty calculated using the two-point geometric correlation of local stiffnesses. \emph{Inset} shows the two-point geometric correlation $\rho_{ij}$ for local stiffness, defined as the covariance between log-local stiffnesses $\log k_{\mathrm{loc}}^{(i)}$ and $\log k_{\mathrm{loc}}^{(j)}$ divided by the logarithm of the geometric standard deviation of local stiffness squared $(\log \sigma_{\mathrm{loc}})^2$, versus distance $R_{ij}$. (\emph{B}) Fractional uncertainty of stiffness inference from three measurements, defined as the geometric standard deviation $\sigma_{\mathrm{3}}$ of the geometric mean of a random sample of $N=3$ local stiffnesses measured by force dipoles whose centers lie within spheres (\emph{Inset}: blue) and prolate spheroids of equivalent volume (\emph{Inset}: green, aspect ratio 2:1, and red, aspect ratio 3:1) versus radius of spheres. Dashed black line shows the geometric standard deviation $\sigma_{\mathrm{loc}}^{1/\sqrt{3}}$ of the geometric mean of $N=3$ independent measurements.
}
\end{figure*}

\newpage

\end{document}